\documentclass[sigconf]{acmart}
\usepackage{amsmath}
\usepackage{natbib}
\usepackage{float}
\usepackage{graphicx}
\usepackage{epstopdf}
\usepackage{subfigure}
\usepackage{amsthm}
\usepackage{enumerate}
\usepackage{multirow}
\usepackage[normalem]{ulem}
\usepackage[ruled,linesnumbered]{algorithm2e}  
\usepackage{algorithmic}
\usepackage{xcolor} % 引入xcolor包
\usepackage{bbm}
\usepackage{bm}
\useunder{\uline}{\ul}{}
\graphicspath{{./fig/}}
\newtheorem{definition}{Definition}

\newtheorem{theorem}{Theorem}

\usepackage{balance}
\usepackage{hyperref}
\hypersetup{
    colorlinks=true, % 启用颜色链接
    urlcolor=blue,    % 定义URL链接的颜色
    linkcolor=blue,   % 定义内部链接的颜色（例如目录）
    citecolor=blue    % 定义引用的颜色
}

\AtBeginDocument{%
  \providecommand\BibTeX{{%
    \normalfont B\kern-0.5em{\scshape i\kern-0.25em b}\kern-0.8em\TeX}}}

\setcopyright{acmcopyright}

\copyrightyear{2024}
\acmYear{2024}
\setcopyright{acmlicensed}\acmConference[KDD '24]{Proceedings of the 30th
ACM SIGKDD Conference on Knowledge Discovery and Data Mining}{August
25--29, 2024}{Barcelona, Spain}
\acmBooktitle{Proceedings of the 30th ACM SIGKDD Conference on Knowledge
Discovery and Data Mining (KDD '24), August 25--29, 2024, Barcelona, Spain}
\acmDOI{10.1145/3637528.3671817}
\acmISBN{979-8-4007-0490-1/24/08}

\begin{document}

%%
%% The "title" command has an optional parameter,
%% allowing the author to define a "short title" to be used in page headers.
% \title{A user watch model for video recommendation}
% \title{Counterfactual Watch Time Modification for Video Recommendation}
\title{Counteracting Duration Bias in Video Recommendation via Counterfactual Watch Time}
% \setCJKfamilyfont{font01}{AR PL KaitiM GB}
%%
%% The "author" command and its associated commands are used to define
%% the authors and their affiliations.
%% Of note is the shared affiliation of the first two authors, and the
%% "authornote" and "authornotemark" commands
%% used to denote shared contribution to the research.
% \author{Ben Trovato}
% \authornote{Both authors contributed equally to this research.}
% \email{trovato@corporation.com}
% \orcid{1234-5678-9012}
% \author{G.K.M. Tobin}
% \authornotemark[1]
% \email{webmaster@marysville-ohio.com}
% \affiliation{%
%   \institution{Institute for Clarity in Documentation}
%   \streetaddress{P.O. Box 1212}
%   \city{Dublin}
%   \state{Ohio}
%   \country{USA}
%   \postcode{43017-6221}
% }

\author{Haiyuan Zhao}
\authornote{Both authors contributed equally to this research.}
\affiliation{
\institution{School of Information\\Renmin University of China}
\city{Beijing}\country{China}
}
\email{haiyuanzhao@ruc.edu.cn}

\author{Guohao Cai}
\authornotemark[1]
\affiliation{%
  \institution{Huawei Noah’s Ark Lab}
\city{Shenzhen}\country{China}  }
\email{caiguohao1@huawei.com}

\author{Jieming Zhu}
\affiliation{%
  \institution{Huawei Noah’s Ark Lab}
\city{Shenzhen}\country{China}  }
\email{jiemingzhu@ieee.org}

\author{Zhenhua Dong}
\affiliation{%
  \institution{Huawei Noah’s Ark Lab}
\city{Shenzhen}\country{China}
  }
\email{dongzhenhua@huawei.com}

\author{Jun Xu}
\authornote{Jun Xu is the corresponding author. Work partially done at Engineering Research Center
of Next-Generation Intelligent Search and Recommendation, Ministry of Education.
}
\affiliation{%
  \institution{\mbox{Gaoling School of Artificial Intelligence}\\ Renmin University of China}
\city{Beijing}\country{China}
  }
\email{junxu@ruc.edu.cn}

\author{Ji-Rong Wen}
\affiliation{%
  \institution{\mbox{Gaoling School of Artificial Intelligence}\\ Renmin University of China}
\city{Beijing}\country{China}
  }
\email{jrwen@ruc.edu.cn}

%%
%% By default, the full list of authors will be used in the page
%% headers. Often, this list is too long, and will overlap
%% other information printed in the page headers. This command allows
%% the author to define a more concise list
%% of authors' names for this purpose.
% \renewcommand{\shortauthors}{Trovato and Tobin, et al.}
\renewcommand{\shortauthors}{Haiyuan Zhao et al.}

%%
%% The abstract is a short summary of the work to be presented in the
%% article.
\begin{abstract}

In video recommendation, an ongoing effort is to satisfy users' personalized information needs by leveraging their logged watch time. However, watch time prediction suffers from duration bias, hindering its ability to reflect users' interests accurately. Existing label-correction approaches attempt to uncover user interests through grouping and normalizing observed watch time according to video duration. Although effective to some extent, we found that these approaches regard completely played records (i.e., a user watches the entire video) as equally high interest, which deviates from what we observed on real datasets: users have varied explicit feedback proportion when completely playing videos. In this paper, we introduce the \emph{counterfactual watch time}~(CWT), the potential watch time a user would spend on the video if its duration is sufficiently long. Analysis shows that the duration bias is caused by the truncation of CWT due to the video duration limitation, which usually occurs on those completely played records. Besides, a Counterfactual Watch Model (CWM) is proposed, revealing that CWT equals the time users get the maximum benefit from video recommender systems. Moreover, a cost-based transform function is defined to transform the CWT into the estimation of user interest, and the model can be learned by optimizing a counterfactual likelihood function defined over observed user watch times. Extensive experiments on three real video recommendation datasets and online A/B testing demonstrated that CWM effectively enhanced video recommendation accuracy and counteracted the duration bias.

\end{abstract}

%%
%% The code below is generated by the tool at http://dl.acm.org/ccs.cfm.
%% Please copy and paste the code instead of the example below.
%%

%%
%% Keywords. The author(s) should pick words that accurately describe
%% the work being presented. Separate the keywords with commas.
%\keywords{learning to rank, examination bias, trust bias, joint learning}

%% A "teaser" image appears between the author and affiliation
%% information and the body of the document, and typically spans the
%% page.
% \begin{teaserfigure}
%   \includegraphics[width=\textwidth]{sampleteaser}
%   \caption{Seattle Mariners at Spring Training, 2010.}
%   \Description{Enjoying the baseball game from the third-base
%   seats. Ichiro Suzuki preparing to bat.}
%   \label{fig:teaser}
% \end{teaserfigure}

%%
%% This command processes the author and affiliation and title
%% information and builds the first part of the formatted document.
\maketitle

\section{Introduction}
The rising of video content platforms have attracted billions of users and become more frequent in the daily use of users nowadays~\citep{Covington2016Deep, Davidson2010YouTube, Liu2019User, Liu2021Concept}. To satisfy users' information needs and enhance their engagement, developing accurate and personalized video recommender systems is critical. It is essential to incorporate various feedback signals that reflect users' interests to achieve this goal. 
In the video scenario, watch time has been commonly employed as an indicator of user interest and can be leveraged to enhance the accuracy of video recommender systems~\citep{Covington2016Deep,DBLP:conf/www/0001XZX0ZWZXZJG23}. 

%Research efforts have been undertaken to improve by leveraging the watch time~\citep{}. 

Like other implicit feedback signals (e.g., user click) in recommender systems, directly inferring the user interest labels from the watch time is also hurt by the bias problem. One crucial bias is duration bias; that is, users' watch time is not only related to their interest but also affected by the duration (length) of the video~\citep{Zhan2022Deconfounding,Zheng2022DVR}. It has been observed that users naturally tend to watch for more time on longer videos, making watch time no longer a faithful reflection of the user's interest.
%The duration bias refers to the tendency of users to watch more time for longer videos, irrespective of their inherent interest. The existence of duration bias makes watch time no longer a faithful reflection of user interest.
% , thereby engendering biased recommendation models when fitting it.

% \begin{figure}
%     \subfigure[Explicit positive feedback ratio]{
%     \includegraphics[width=0.2\textwidth]{figs/kuairand_posi_ratio_bar.pdf}}
%     \quad
%     \subfigure[Explicit negative feedback ratio]{
%     \includegraphics[width=0.2\textwidth]{figs/kuairand_neg_ratio_bar.pdf}}
%     \caption{Users' explicit feedback ratio in completely played
%     records of the KuaiRand dataset. (a) The positive explicit feed-
%     back ratio grouped by duration. (b) The negative explicit
%     feedback ratio grouped by duration.}
%     \label{fig: behavior confusion}
% \end{figure}

\begin{figure}
    \subfigure[KuaiRand]{
    \includegraphics[width=0.22\textwidth]{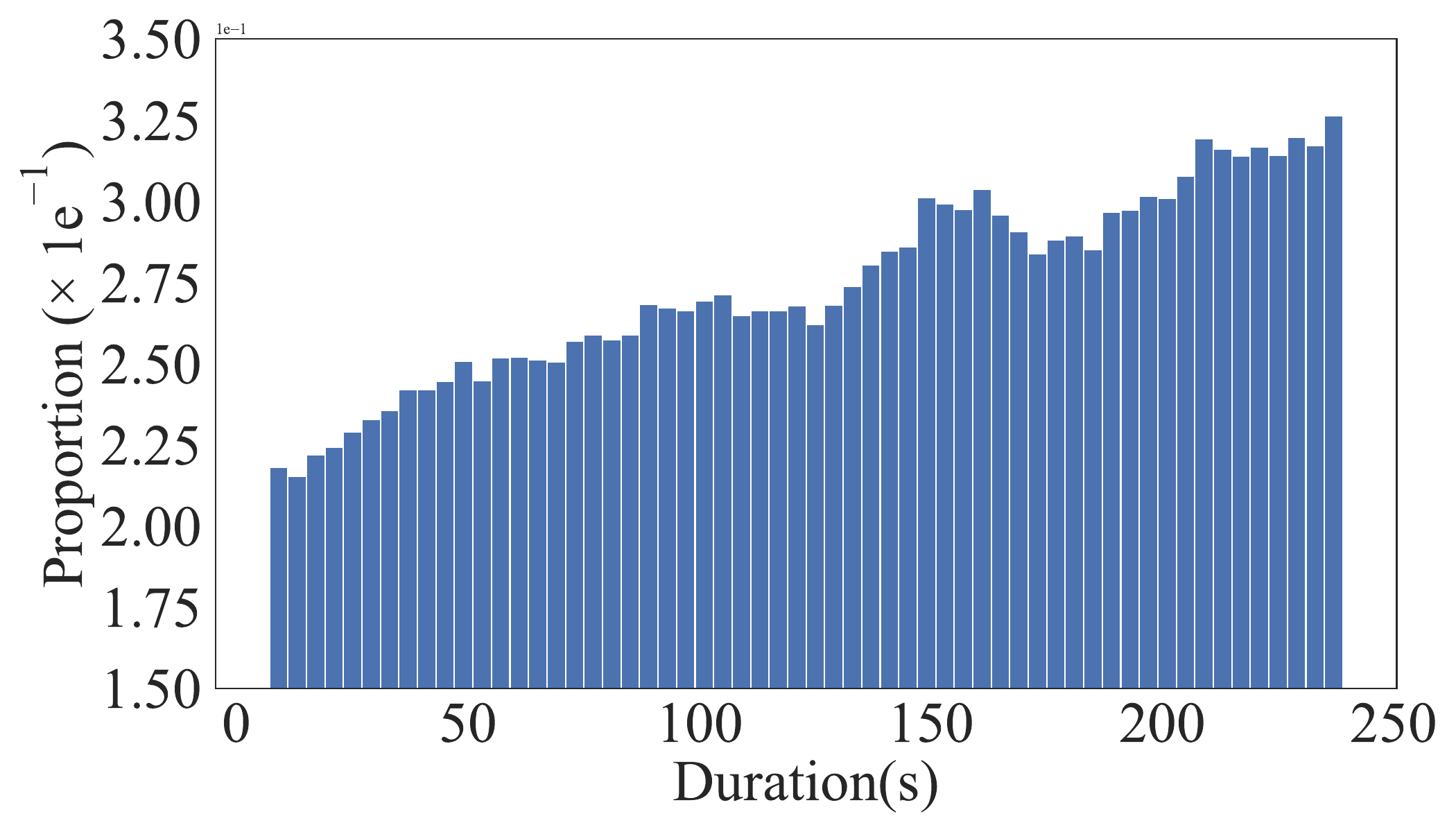}}
    \quad
    \subfigure[WeChat]{
    \includegraphics[width=0.22\textwidth]{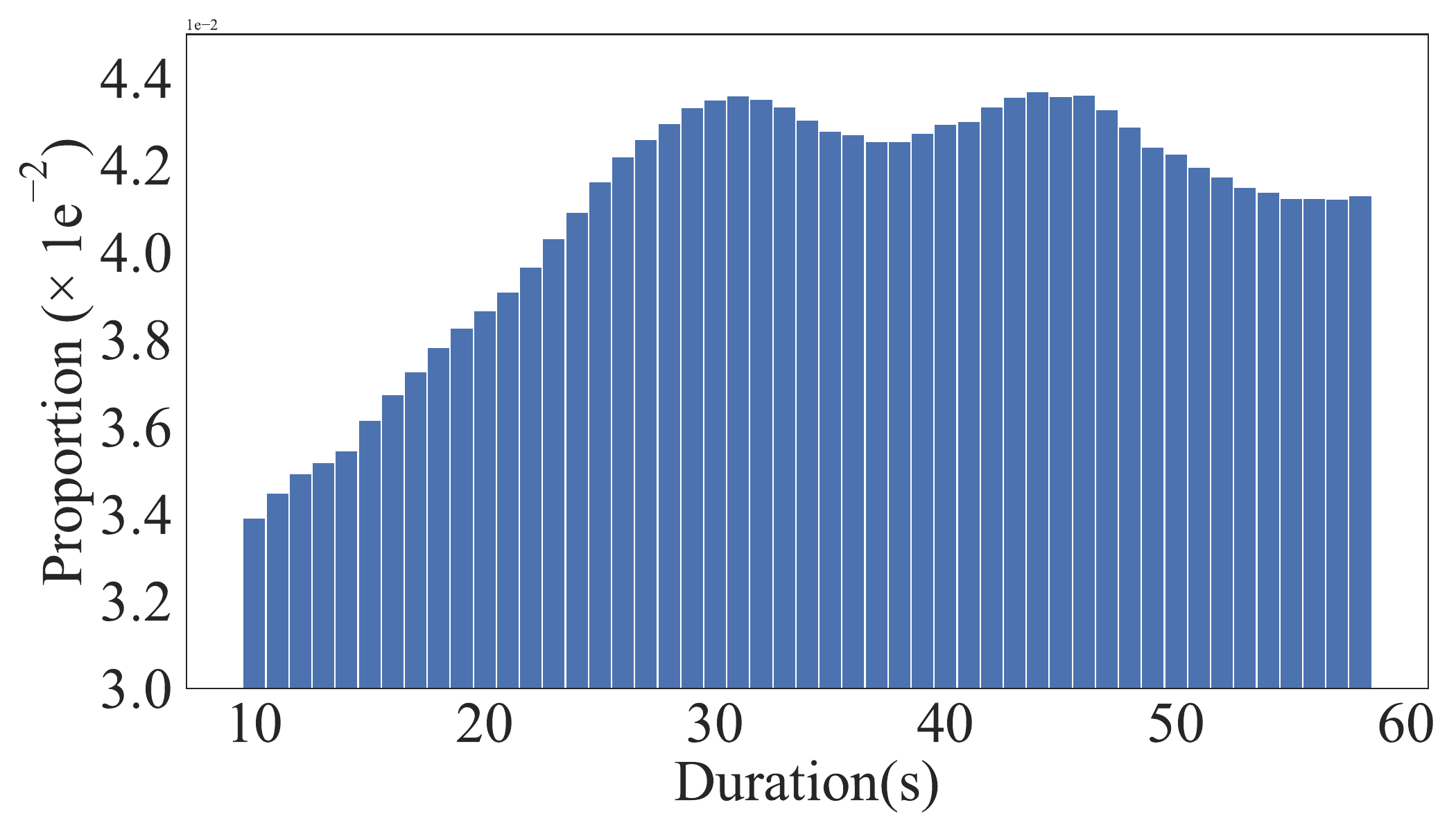}}
    \caption{Users' explicit feedback proportion in completely played records grouped by video duration of (a) KuaiRand dataset (b) WeChat dataset.}
    \label{fig: behavior confusion}
\end{figure}

\begin{figure}[t]
    \includegraphics[width=0.48\textwidth]{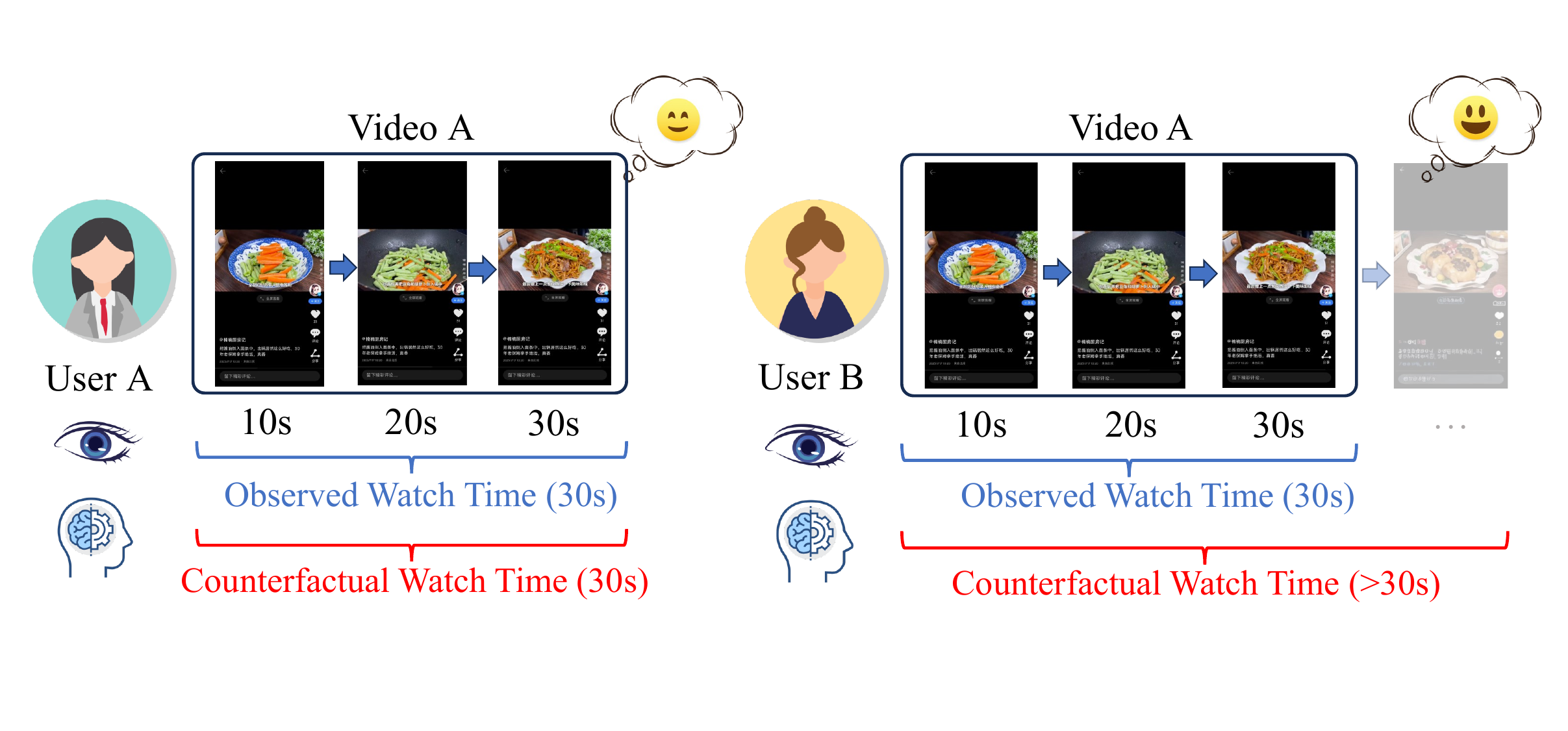}
    \caption{Comparison between counterfactual watch time and observed watch time. Users A and B have the same observed watch time but different counterfactual watch times.}
    % Left: user actively stop watching. Right: user passively stops watching due to the limit of duration.
    \label{fig:ct_watch}
\end{figure}

Existing approaches like Play Completion Rate (PCR), Watch Time Gain (WTG)~\citep{Zhan2022Deconfounding} and Quantile-based method (D2Q)~\citep{Zheng2022DVR} regard duration bias as a problem of inconsistent watch time scales caused by the video duration. The longer the video duration, the larger the watch time scale, resulting in a longer average watch time. Therefore, these methods group and normalize the watch time according to the video duration, which keeps the watch time corresponding to different video duration consistent in scale. For example, PCR directly divides the watch time according to the corresponding video duration and then normalizes the watch time into intervals of 0 to 1. These normalized values are treated as estimated labels of user interest in downstream tasks. Since these methods mitigating bias via correcting labels, they can be categorised as \textbf{label-correction methods}. 
In practice, both video ranking based on user interest and accurately predicting watch time are crucial tasks. The advantage of the label-correction method is that it can not only provide an unbiased estimate of the user interest score but also predict watch time through the inverse transformation of the normalization function. In contrast, other existing methods, such as the feature-based methods: DCR~\citep{He2023Confounding} and CVRDD~\citep{Tang2023Video}, primarily focus on the unbiased estimation of user interest scores. While these methods have proven effective, they struggle to simultaneously estimate both watch time and user interest scores as effectively as label-correction methods.

Although existing label-correction methods have shown effectiveness in some
extent, we argue that there are still limitations. In existing methods, after the normalization, the completely played records (i.e., the user watched the whole video) are usually treated as the highest user interest, regardless of the video duration. Typically, records with explicit feedback can reflect user engagement and interest to some extent. However, from the explicit user feedback signals provided in KuaiRand dataset\footnote{\url{https://kuairand.com/}} and WeChat dataset \footnote{\url{https://algo.weixin.qq.com/}}, we observed that user explicit feedback in these completely played records does not align well with current methods. Specifically, Fig.~\ref{fig: behavior confusion} shows the proportion of explicit positive feedback(i.e., like and forward) received by those completely played records, grouped by the video duration. 
In existing methods~\citep{Zhan2022Deconfounding,Zheng2022DVR}, since the user has fully watched the video, these completely played records should be considered as indicating equally high interest. However, Fig.~\ref{fig: behavior confusion} clearly shows that, on both two datasets, longer completely played videos have a higher proportion of positive feedback, even though the short videos are also completely played. 

% Similarly, Fig.~\ref{fig: behavior confusion}(b) shows the proportion of the completely played records while receiving explicit negative feedback (i.e., hate), grouped by the video duration. Also, the proportion was expected to be nearly zero for all duration because all the records are considered to have the highest user interest. However, Fig.~\ref{fig: behavior confusion}(b) showed that negative feedback is present in almost all duration completely played videos and is higher in shorter completely played videos.

% The results indicate that users have different interest levels when they play videos completely. 
The results in Fig.~\ref{fig: behavior confusion} suggest that even if users have watched the entire video, these completely played records may reflect varying levels of interest. 
Moreover, the shorter the video duration, the lower the user interest level that a completely played record can represent, as it corresponds to a lower explicit feedback proportion.
This inspires us to derive a novel interpretation of the duration bias: \emph{There exists a potential watch time that faithfully reflects user interest, which is truncated by the video's duration, thereby resulting in a duration bias.}. As illustrated in Figure~\ref{fig:ct_watch}, let's consider two users, A and B, engaged in watching the same video with a 30s duration. Both users have completely played the video, and their watch times are recorded as 30s in the log data. However, after watching the same video, User B exhibits a higher level of interest than User A, thus appearing to be left wanting more. We then consider a counterfactual question: \emph{How long will the user watch if the video duration is sufficiently long?} It becomes clear that User A is less likely, or even unlikely, to extend her observed watch time. The logged 30s watch time for user A can sufficiently reflect her interest. Conversely, User B prefers exceeding her observed watch time. Since her information needs are not fully satisfied after watching this video, the logged 30s watch time cannot represent B's interest. 

To this end, we argue that there exists a \textbf{counterfactual watch time}~(CWT) corresponding to user interest, which is partially observed due to the truncation by video duration. For incompletely played records, the CWT equals its observed watch time, thus reflecting user interest. For completely played records, the CWT could be longer than the observed watch time due to the truncation by video duration. The user's true interest level could be higher than that directly inferred from the observed watch time. 

% User A discontinues playing when reaching the video's duration, while User B completes playing but expresses a desire to continue. Both users watched 30 seconds but had very different interest levels. We consider a counterfactual question: \emph{How long will the user watch if the video duration were extended?} it becomes evident that users who actively stop watching (e.g., User A) would not extend their observed watch time. Conversely, the users who passively stop watching (e.g., User B) prefer exceeding their observed watch time. To this end, we argue that there exists a \textbf{counterfactual watch time} corresponding to user interest. 
% % \jm{provide a definition of counterfactual watch time here} 
% For incompletely played records, the observed watch time equals its counterfactual watch time, thus can reflecting users' interest. For completely played records, the observed watch time may be shorter than the counterfactual watch time due to the truncation of video duration. The user's true interest level may be higher than that directly inferred from the observed watch time. 

The CWT can explain the results in Fig.~\ref{fig: behavior confusion} well: various CWT are truncated by video duration in an entirely played video. Therefore, the same observed watch time may correspond to various CWT and thus cannot distinguish user interest level. Furthermore, the completion of shorter videos occurs more easily than longer ones, exacerbating the truncation of CWT. This results in a more pronounced duration bias, impacting the accuracy of interest measurement based on watch time.

To model the CWT and estimate user interests, we propose a Counterfactual Watch Model (CWM) that adopts an economic perspective to model users' CWT. Specifically, CWM treats user watching as a process of accumulating watching rewards, where the marginal rewards are indicative of user interest, and the invested watch time corresponds to the user watching cost. At the time point where a user's marginal cost equals the marginal rewards, the user attains the maximum cumulative benefit, making her actively stop watching. This time point corresponds to the aforementioned CWT. Then, a cost-based transform function is derived to transform the CWT to the estimated user interest. The duration-debiased recommendation model can be learned by optimizing a counterfactual likelihood function defined over observed user watch times. In summary, CWM attempts to model users' consumption behavior, specifically in video scenarios. Similar to the commonly used click model~\citep{Chuklin2016Click}, our CWM is beneficial for both relevance ranking and watch time prediction via effectively modeling user behaviors.

The major contributions of this work are:
\begin{enumerate}[(1)]
\item We provide a novel concept called \emph{counterfactual watch time} (CWT) for interpreting the essence of duration bias;
\item We propose a method named CWM for modeling CWT. We further develop a cost-based transform function and counterfactual likelihood function for learning a duration-debiased recommendation model;
\item We conduct experiments on three real video recommendation datasets and online A/B testing. The result improvements demonstrate the effectiveness of our CWM.
\end{enumerate}

\section{Related work}
\textbf{Video Recommendation}. In the evolution from traditional video recommendation to TV show, various methods have been developed to enhance the user experience and accuracy of recommendations. ~\citet{park2017rectime} introduced a system that incorporates time factors and user preferences using 4-dimensional tensor factorization to improve recommendation accuracy. ~\citet{cho2019no} presented a recommendation method that accounts for user feedback within specific watchable intervals to enhance user satisfaction with TV show recommendations. ~\citet{qin2023learning} proposed a model that identifies and adapts to the behavior of multiple users interacting with a TV system, thereby improving the personalization of recommendations. Most research works have transferred from traditional video to micro-video scenarios in the mobile internet era. ~\citet{WLR} introduced the funnel architecture of YouTube recommender system. They predicted the expected watch time from training samples with weighted logistic regression, which utilizes observed watch time as the weight of positive samples' loss. Multi-task methods~\cite{MMOE,PLE,PEPNet,DBLP:conf/recsys/ZhaoHWCNAKSYC19} have been proposed to improve metrics such as watch time prediction, relevance of user-item pair, and number of video views together.~\citet{MMOE} extend the mixture-of-experts~\cite{MOE} architecture to multi-gate expert knowledge integration.~\citet{PLE} proposed a shared learning structure to address the seesaw phenomenon.~\citet{PEPNet} proposed a plug-and-play parameter and embedding
personalized network for a multi-domain and multi-task recommendation. 
% In another aspect, methods based on reinforcement learning have also been deployed in some industrial systems. Chen et al.~\cite{DBLP:conf/wsdm/ChenBCJBC19} scaled REINFORCE to a production recommender system with an action space on the orders of millions for the first time. Cai et al.~\cite{DBLP:conf/www/0001XZX0ZWZXZJG23} formulated the micro-video recommendation problem as a constrained Markov decision process, and proposed a two-stage constrained actor-critic method to optimize the main goal watch time and other auxiliary goals. 
% such as Like, Follow, Share user interactions.

\textbf{Counterfactual Information Retrieval}. Most information retrieval systems consider users' implicit feedback as a supervision signal to infer their true interests. However, implicit feedback is influenced not only by users' interests but also by external factors. Consequently, user interest signals are often concealed within implicit feedback and remain unobserved. Researchers have drawn inspiration from causal inference techniques~\citep{Wu2022survey} and developed counterfactual information retrieval technology to address the biases inherent in implicit feedback. Previous work in counterfactual IR primarily focuses on mitigating position bias~\citep{Joachims2017Unbiased,Ai2018Unbiased,Yuan2020Unbiased,Chen2021Adapting}, popularity bias~\citep{Zhang2021Causal,Zheng2021Disentangling,Wei2021Model} , and selection bias~\citep{schnabel2016recommendations,Wang2016Learning,Saito2020Unbiased}.
% Inspired by causal inference~\citep{Wu2022survey}, a large number of debiasing methods are proposed for mitigating biases above, which includes propensity-based methods~\citep{Joachims2017Unbiased,Zhang2022Counteracting}, backdoor adjustment methods~\citep{Wei2021Model,Zhang2021Causal} and causal embedding methods~\citep{Zheng2021Disentangling,Chen2021Adapting,Bonner2018Causal}.
However, duration bias becomes a crucial concern when it comes to video recommendation, which has been discussed in existing studies~\citep{Zhan2022Deconfounding,Zheng2022DVR, Quan2023Video, Tang2023Video, He2023Confounding}. In contrast to our approach, current methods for correcting duration bias cannot effectively explain and eliminate duration bias.

\textbf{Click model in Information Retrieval}. Modeling user behaviors plays a vital role in enhancing the performance of information retrieval systems. The ability to accurately model user behaviors allows a retrieval system better to fulfill users' information needs~\citep{Chuklin2016Click}. To this end, many models have been proposed to explain or predict user click behavior in various contexts: cascade model (CM)~\citep{Craswell2008Experimental}, user browsing model (UBM)~\citep{Dupret2008User} and dynamic Bayesian network (DBN) model~\citep{Chapelle2009Dynamic} model users' click behavior in desktop searching with different assumption; mobile click model (MCM)~\citep{Mao2018Constructing} and F-shape Click Model (FSCM)~\citep{Fu2023Click} further extend the understanding of users' click behaviors on mobile devices. Moreover,~\citet{Borisov2016Neural} and~\citet{Chen2020Context} develop the click model into neural networks, which enable automatic dependency detection. Unlike the above click models, the UWM proposed in this paper focuses on modeling and explaining users' watching behavior since it is a better quantitative indicator of user preferences in video feeds.

\section{Counterfactual watch time}
In this section, we will first define the video recommendation problem and the counterfactual watch time~(CWT); then we will provide supporting evidences for our proposed CWT. We also present an economic view of the user watch behavior based on the watch cost and reward. Finally, we point out the current methods' limitation from the CWT viewpoint.

\subsection{Definition of counterfactual watch time}
\label{sec: ct_wt define}
The task of user interest and watch time prediction can be formalized as follows: Suppose $\mathcal{U}$ and $\mathcal{V}$ are the sets of users and videos, respectively. We can record user $u\in\mathcal{U}$'s watching behavior on video $v\in\mathcal{V}$ as $\mathcal{D}_{u,v}=\{\mathbf{x}_{u,v}, w_{u,v}, d_{v}\}$, where $\mathbf{x}_{u,v}\in\mathbb{R}^k$ represents the feature vector of the sample pair and $k$ is the feature dimension. $w_{u,v}\in \mathbb{R}^+$ denotes user $u$'s observed watch time on video $v$ (e.g., in seconds), while $d_{v}\in \mathbb{R}^+$ is the duration of video $v$. 
%Based on a set of collected log data $\mathcal{D} =\{\mathcal{D}_{u,v}: u\in\mathcal{U}, v\in\mathcal{V}\}$, a prediction model is learned which aims to estimate the user interest and the user watch time.

Next, we will introduce a novel concept called \emph{counterfactual watch time}~{CWT}, which is denoted as $w^{c}_{u,v}$. As we have discussed before, the CWT can be defined as: 
\begin{definition}[counterfactual watch time]
\label{def: counterfactual watch time}
    For user $u$ and video $v$, the CWT $w^{c}_{u,v}\in \mathbb{R}$ is defined as the time users want to watch based on the user's interest $r_{u,v}$ if the video duration is sufficiently long. There is no correlation between $w^{c}_{u,v}$ and video duration $d_{v}$.
% \[
%      w_{u,v} = \mathrm{min}(w^{c}_{u,v}, d_{v})
% \]
\end{definition}

The CWT $w^{c}_{u,v}$ corresponds to user interests.
% , where $g(\cdot)$ is transform function between $w^{c}_{u,v}$ and $r_{u,v}$.
However, CWT $w^{c}_{u,v}$ does not always equal the observed watch time $w_{u,v}$ since it can be truncated by video duration $d_{v}$ in practice\footnote{CWT may also be truncated at 0, making it difficult to discern how much a user dislikes the video. However, this study is more concerned with the videos users like than those they dislike.}. Their relationship is formulated as follows:
\begin{equation} 
\label{eq: cwt_and_awt}
    \begin{split}
        w_{u,v} = \mathrm{min}(w^{c}_{u,v}, d_{v}) \Longleftrightarrow \left\{
        \begin{aligned}
            & w^{c}_{u,v}= w_{u,v},\quad \mathrm{if}~~ w_{u,v}<d_{v} \mathrm{;} \\
            & w^{c}_{u,v}\geq w_{u,v},\quad \mathrm{if}~~ w_{u,v}=d_{v} \mathrm{;} 
        \end{aligned}\right.
    \end{split}
\end{equation}
Eq.~\eqref{eq: cwt_and_awt} indicate that observed watch time $w_{u,v}$ can be regarded as the truncated variable of CWT $w^{c}_{u,v}$.

\begin{figure}
    \subfigure[KuaiRand]{
    \includegraphics[width=0.22\textwidth]{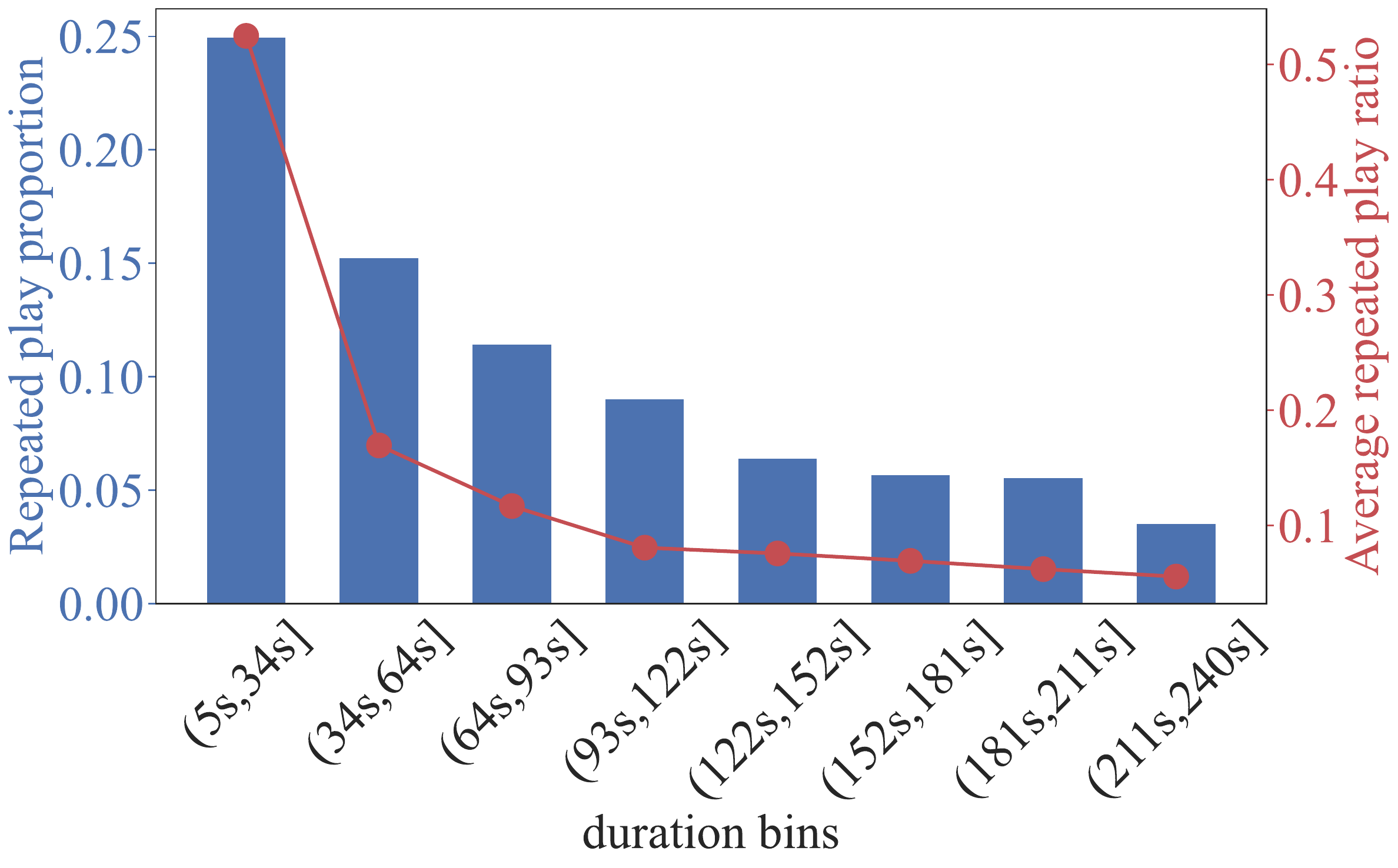}}
    \quad
    \subfigure[WeChat]{
    \includegraphics[width=0.22\textwidth]{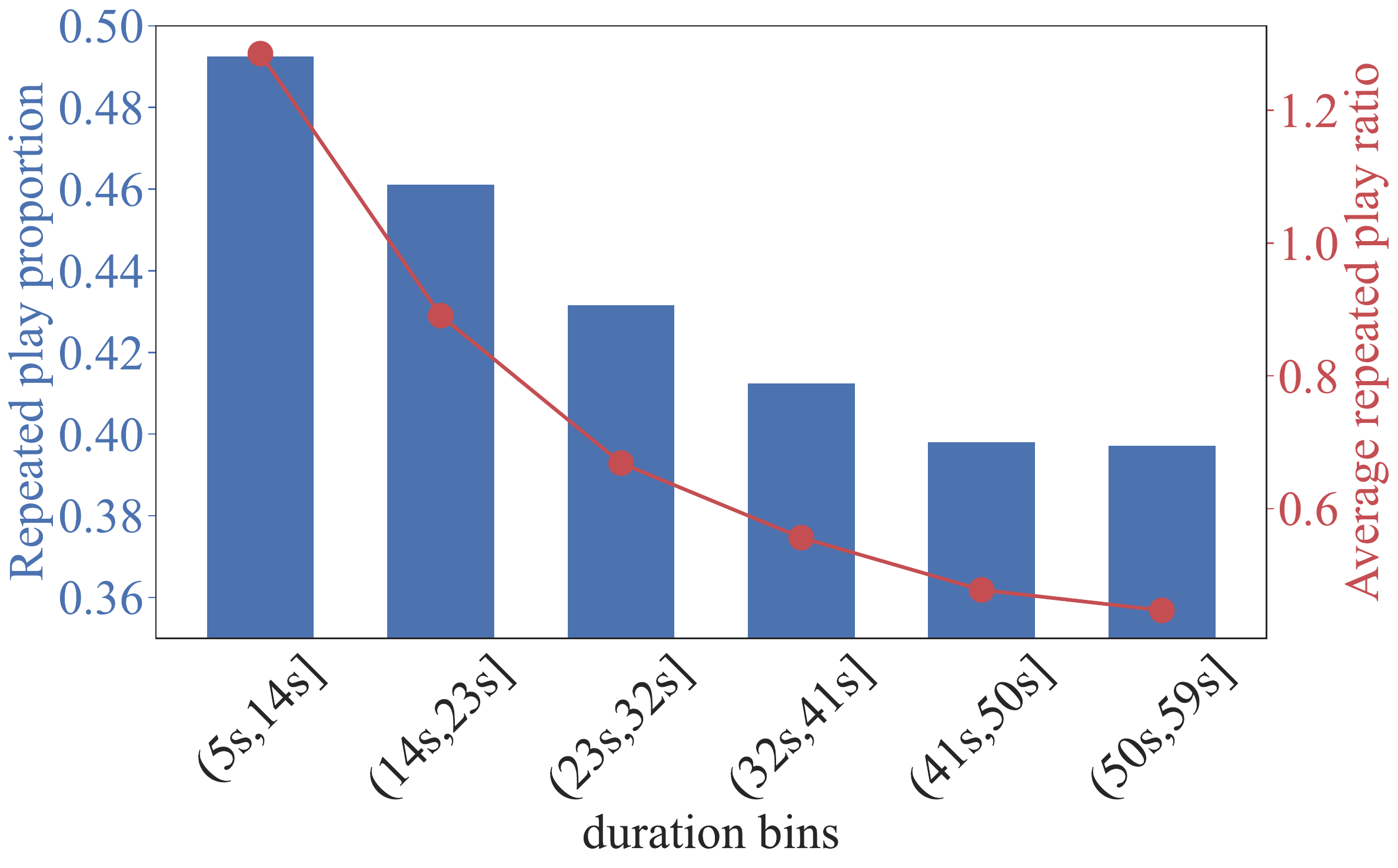}}
    \caption{The repeated play proportion and average repeated play ratio in different duration bins of (a) KuaiRand dataset (b) WeChat dataset.}
    \label{fig: repeat play}
\end{figure}

\begin{figure}
    \subfigure[KuaiRand]{
    \includegraphics[width=0.22\textwidth]{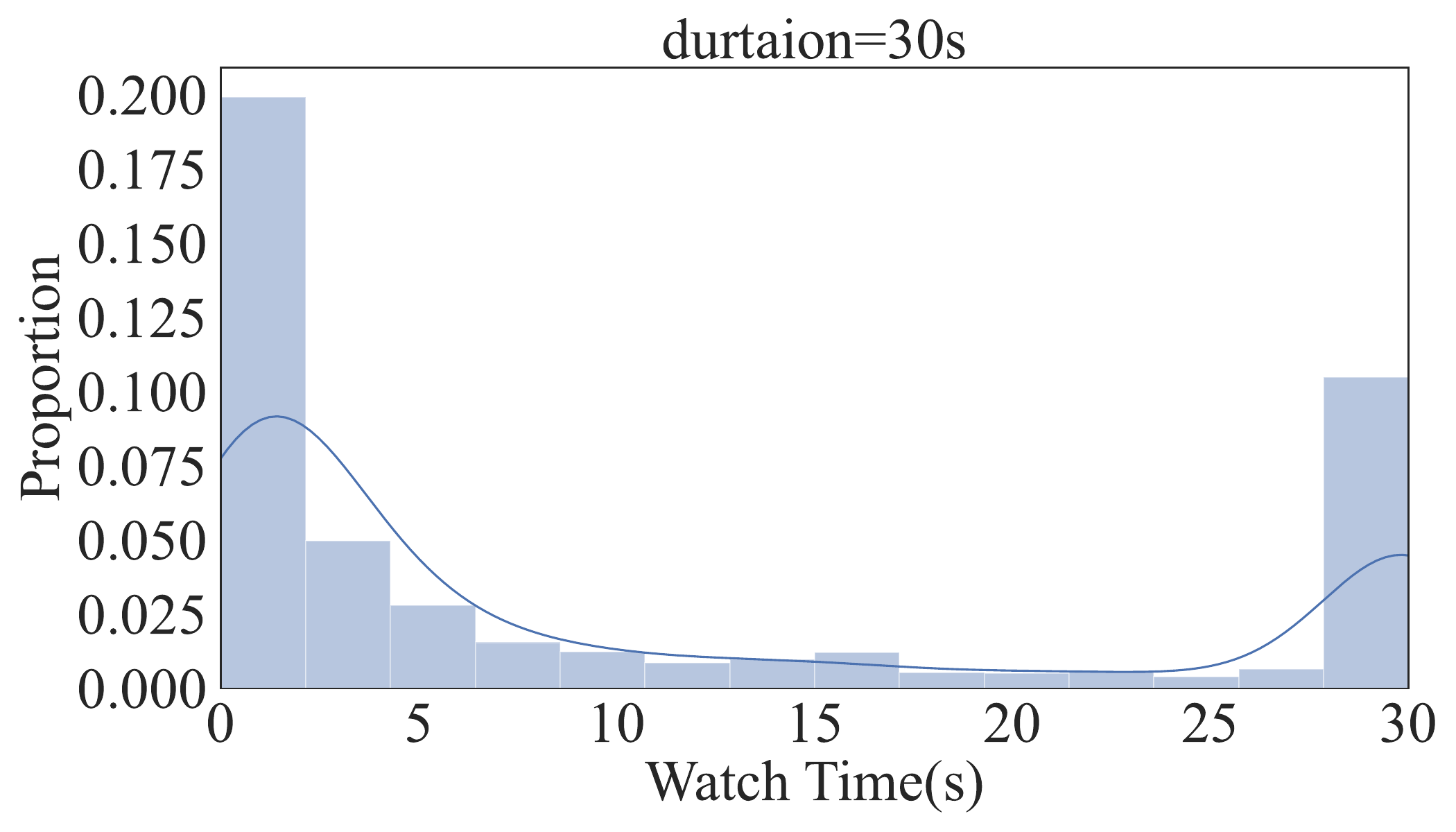}}
    \quad
    \subfigure[WeChat]{
    \includegraphics[width=0.22\textwidth]{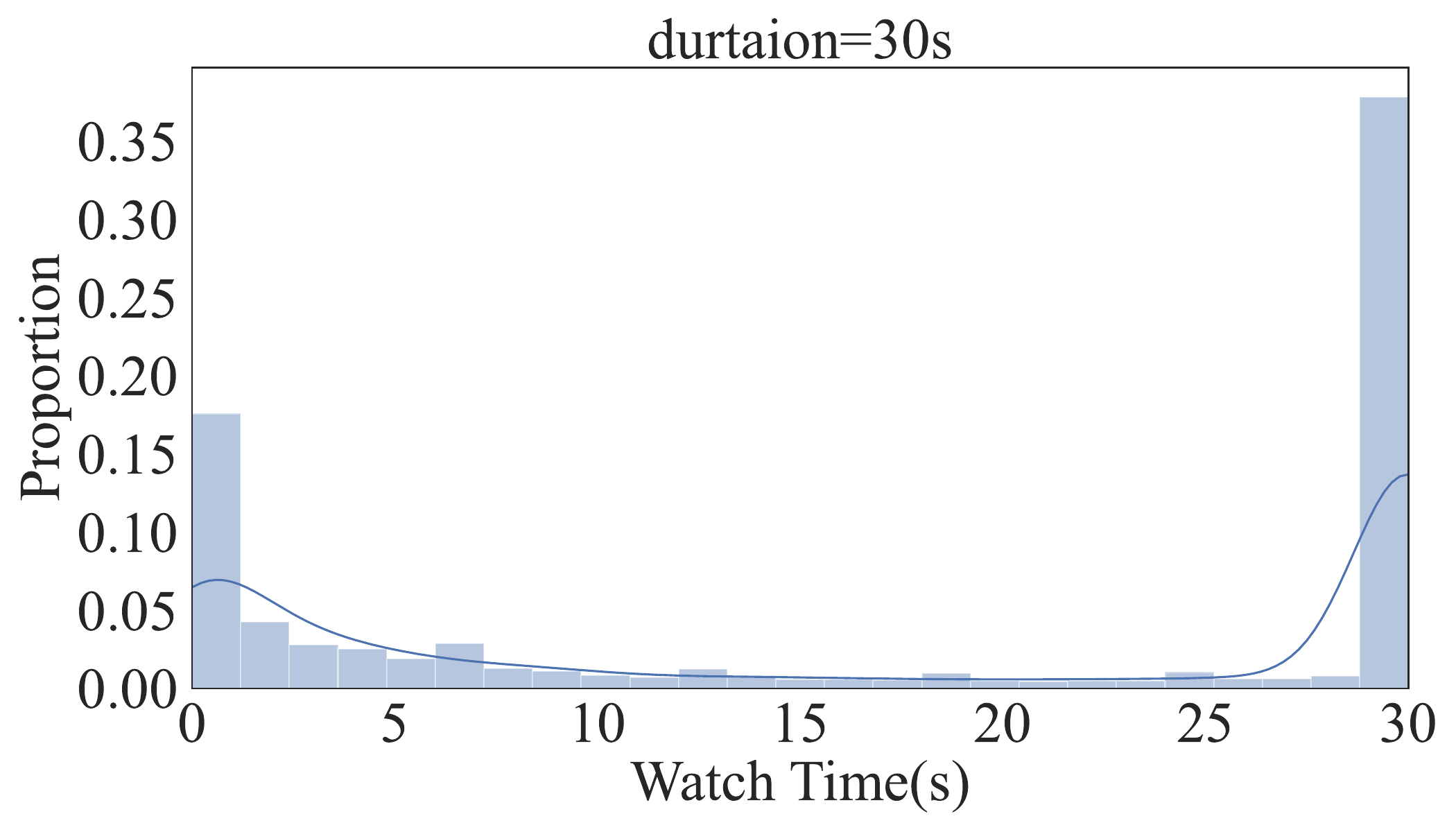}}
    \caption{The bimodal distribution of watch time on (a) KuaiRand dataset (b) WeChat dataset. The video duration is 30s.}
    \label{fig: bimodal dist}
\end{figure}

\subsection{The existence of counterfactual watch time}
Though the CWT is not directly observable from the data, we can still find hints about the existence of CWT in real-world video recommendation datasets. Next, we will use CWT to explain two phenomena presented in real datasets: (i) users' repeated playing and (ii) the bimodal distribution of watch time.

\begin{figure}
    \subfigure[Repeated playing]{
    \includegraphics[width=0.42\textwidth]{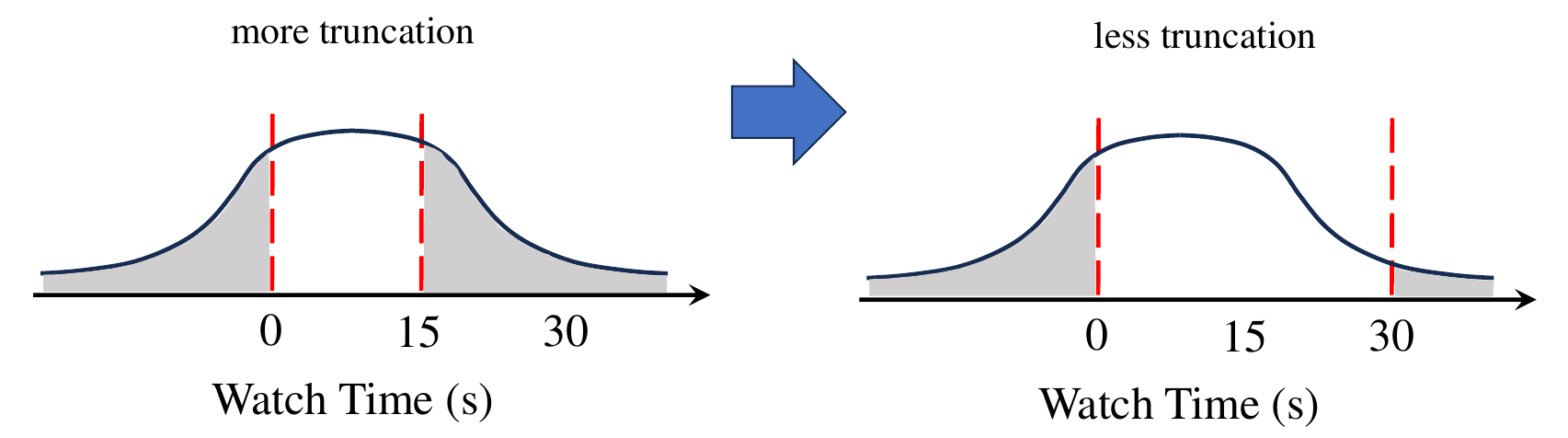}}
    \quad
    \subfigure[Bimodal distribution]{
    \includegraphics[width=0.42\textwidth]{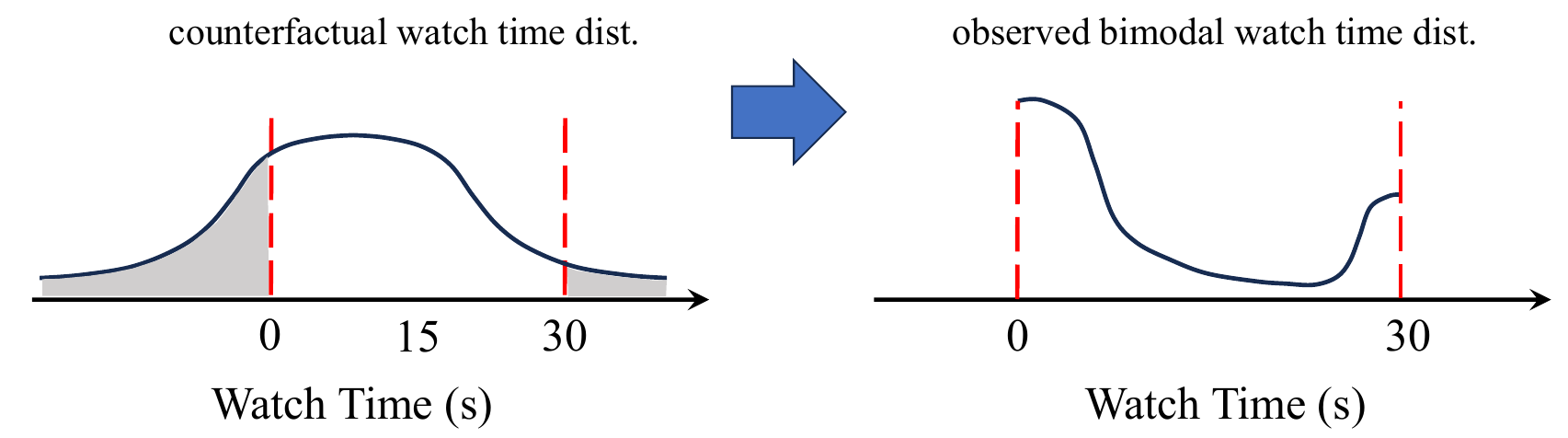}}
    \caption{Employing CWT to explain (a) repeated playing (b) bimodal distribution.}
    \label{fig: bimodal explain}
\end{figure}

% \begin{figure}[t]
%     \includegraphics[width=0.47\textwidth]{figs/bimodal_explain2.pdf}
%     \caption{Explaining the bimodal distribution using counterfactual watch time. }
%     \label{fig: bimodal explain}
% \end{figure}
% The left side of the counterfactual watch time distribution is truncated to 0 and the right side is truncated to duration. Truncated areas are shown in grey.

\subsubsection{Evidence 1: repeated playing}
In real datasets, we found that users may engage in repeated playing (for example, by rewinding the video progress bar), leading to actual watch time that exceeds the video duration. This phenomenon is often due to users' high interest level in the current video. However, the video's duration is insufficient to meet their needs, which is similar to our definition of CWT. Since the definition of CWT necessitates a sufficiently long video duration (Definition~\ref{def: counterfactual watch time}), in this study, we do not equate repeat playing with CWT. In Fig.~\ref{fig: repeat play}, we investigate the repeated playing in both the KuaiRand and WeChat datasets. 
Specifically, we focus on two metrics: (1) \emph{repeated play proportion}, which represents the proportion of repeat played records within the current video duration, and (2) \emph{average repeat play ratio}, which reflects the average extent of repeat playing within the current video duration, defined as $(w_{u,v} - d_{v})/d_{v}$. The results depicted in Fig.~\ref{fig: repeat play} indicate that both the proportion of users' repeat playing and the degree of repeat playing are higher for shorter videos and decrease as video duration increases.

\subsubsection{Evidence 2: bimodal distribution}
Another supporting evidence is the bimodal distribution of users' watch time. For the overall distribution of all the video playing records, existing studies~\citep{Xie2023Reweighting, zhou2018jump} argue that logarithmic watch time obeys the Gaussian distribution. However, when we focus on a given video duration (e.g., 30s), the distribution of observed watch time turns out to be bimodal, as shown in Fig.~\ref{fig: bimodal dist}.
% , which has not only been observed in the public datasets but also been confirmed in our production data. 
The bimodal distribution reveals that most users skip over the recommended video or completely watch it, while only a few users stop watching in the middle of the video playing~\citep{Tang2023Counterfactual,Zhao2023Uncovering}. This abnormal distribution change is less interpreted by existing studies but can be well explained by the CWT.

\subsubsection{Explanation from counterfactual watch time}
Without loss of generality, we assume that CWT $w^{c}_{u,v}$ obeys a Gaussian distribution. However, as we have mentioned in Eq.~\eqref{eq: cwt_and_awt}, all sampled $w^{c}_{u,v}$ will be truncated by duration $d_{v}$. Meanwhile, $w^{c}_{u,v}$ will also be truncated to 0 since all recorded watch times have to be non-negative. The truncated samples are assigned to $d_{v}$ or $0$, respectively. As illustrated in Fig.~\ref{fig: bimodal explain}(a), with the video duration increases, CWT experiences less truncation on the right side, thereby reducing the tendency for users to engage in repeat playing. Meanwhile, in Fig.~\ref{fig: bimodal explain}(b), when the original Gaussian distribution is truncated, it presents a bimodal distribution, as we observed in the real-world dataset. Hence, CWT can successfully interpret the above phenomenon, which in turn supports its existence.  

\subsection{An economic view of user watching}
\label{sec: economic view}

To address the relationship between CWT and user interest, we model the user's watch behavior from an economic perspective. The foundation of CWT within this framework is based on the concepts of \textbf{utility maximization}~\citep{aleskerov2007utility} and \textbf{rational choice}~\citep{vriend1996rational}. This means users decide how much time to allocate to watching videos based on the perceived utility (satisfaction) derived from the content and their resource constraints. 
In accordance with these economic principles, individuals allocate resources—in this case, time—to maximize their utility. Utility can be seen as the \emph{reward} minus the \emph{cost}. For users, the \emph{reward} for watching a video is the information or pleasure derived from the video content, which is determined by their interest to the video. However, users cannot earn unlimited rewards, as they face resource constraints, such as limited time. 
To illustrate the constraint, we introduce the concept of \emph{watching cost}, which refers to the overall effort and resources required to watch a video, including not just the value of time for the user but also the user's mental energy and attention. This \emph{watching cost} highlights that watching a video requires users to allocate their finite resources, which could have been spent on other activities.
% This means the time spent watching a video could have been used for other interesting activities.

In summary, We assume that the user's watching behavior follows the underlying assumptions:
\begin{itemize}
\item \textbf{Diminishing marginal reward}: When the user watches a video, the enjoyment or satisfaction derived per second decreases gradually, which is supported by the habituation phenomenon in the psychology filed~\citep{thompson1966habituation} and existing practice~\citep{Xie2023Reweighting}. The initial marginal watch reward corresponds to user interest.
% The reward for watching a video is the information or pleasure that the user gets from the video, and according to psychological research, users become saturated with repetitive stimuli over a long period of time~\citep{thompson1966habituation}, which leads to a diminishing marginal reward. The marginal watch reward when users start to watch a video corresponds to user interest.
% 
\item \textbf{Constant marginal cost}: Since the watching cost is mainly influenced by context factors, we can mildly assume that context factors are not significantly altered during the period user watching video. Therefore, we consider the marginal watch cost as a constant.
% This assumption indicates that the user pays a constant cost for watching the video per second.
\item \textbf{Rational users}: Users typically act rationally by maximizing their cumulative utility. Users will stop watching the video when the marginal watch cost equals the reward.
\end{itemize}

\begin{figure}
    \subfigure[marginal watch cost/reward]{
    \includegraphics[width=0.22\textwidth]{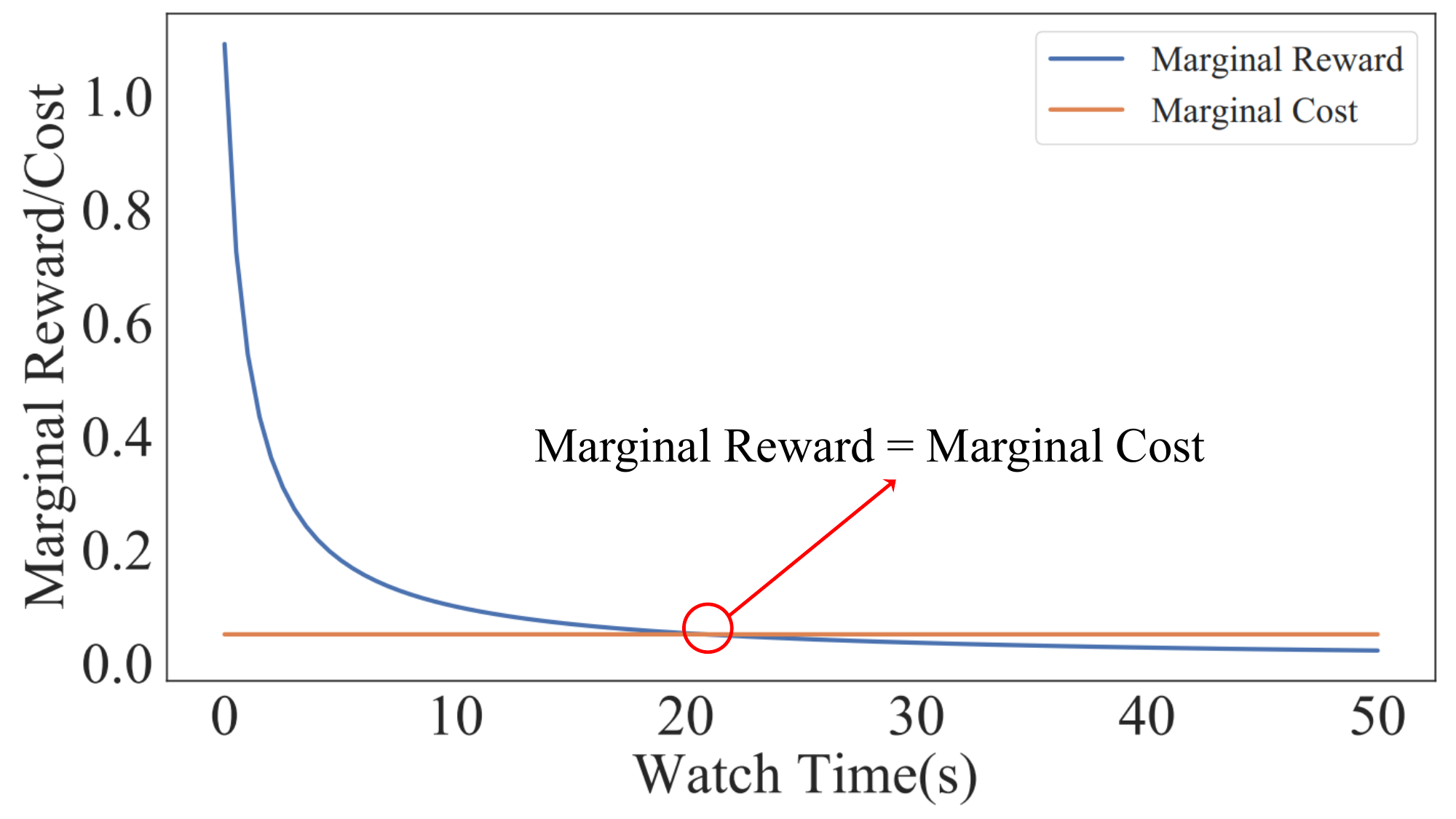}}
    \quad
    \subfigure[cumulative watch cost/reward]{
    \includegraphics[width=0.22\textwidth]{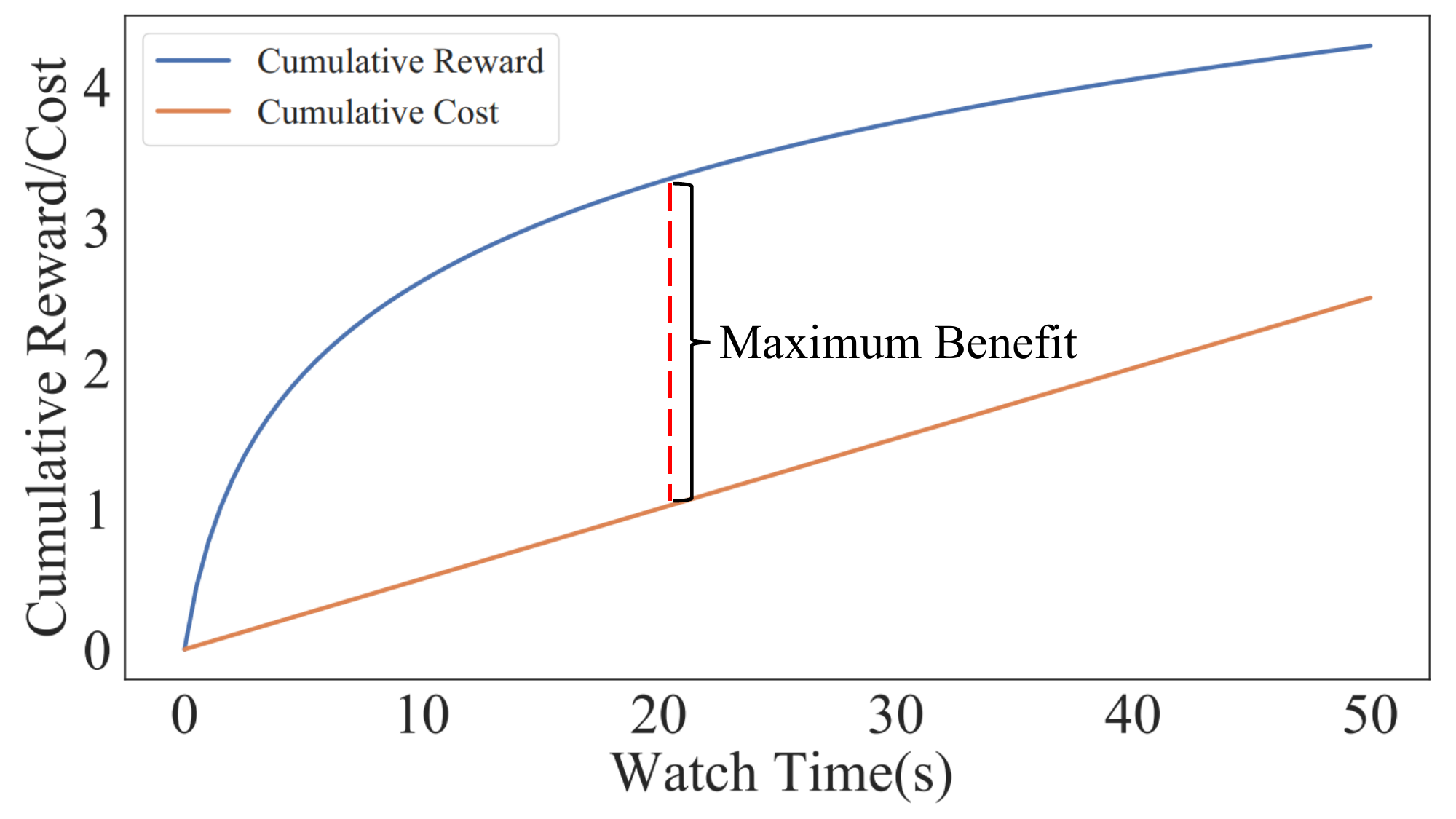}}
    \caption{The economic view of user watching.}
    \label{fig: marginal and cumulative}
\end{figure}

Based on the three assumptions above, we draw the marginal watch cost/reward curves in  Fig.~\ref{fig: marginal and cumulative}(a) and the cumulative watch cost/reward curves in Fig.~\ref{fig: marginal and cumulative}(b). The marginal curves are obtained as derivatives of the cumulative curves. The figure shows that the marginal watch reward decreases monotonically while the marginal watch cost curve is constant. When these two curves meet, a user's marginal watch reward equals the marginal watch cost. At this point, the cumulative utility (i.e., cumulative reward minus cumulative cost) is maximized, as illustrated in Figure~\ref{fig: marginal and cumulative}(b). Formally, we refer to this time point as our proposed CWT. 
% As illustrated in Fig.~\ref{fig: behavior confusion} and Fig.~\ref{fig:ct_watch}, when a completely played video is too short, users' watch time will be truncated too early. It thus cannot reflect users' true interests. 
Our economic view can interpret the phenomenon in Fig.~\ref{fig: behavior confusion} and Fig.~\ref{fig:ct_watch}: users interest level in a video is reflected by the time when they receives their maximum cumulative utility. When the video duration is too short to achieve each user's maximum cumulative utility point, users with either high or low interest will completely play it.

\subsection{Limitation of existing methods}
Finally, we point out the limitation of current methods from the viewpoint of CWT. We will prove that users' true interest cannot be inferred with only a transform function over the observed watch time, as shown in the following theorem:

% \begin{theorem}[cannot infer interest from observed watch time]
\begin{theorem}[observed watch time is not the indicator of interest]
\label{thm: current limit}
% for a given $(u,v)$, the observed watch time is $w_{u,v}\in\mathcal{W}$, user $u$'s interest in video $v$ is $r_{u,v}\in\mathcal{R}$. ~
% $\forall~~\mathcal{W} \subseteq \mathbb{N},~~ \nexists~~ f: \mathcal{W} \to \mathcal{R}$, 
For $\forall~~\mathcal{W} \subseteq \mathbb{R}^+, g\in\mathcal{G}$, given $g: \mathcal{R} \to \mathcal{W}$, we have ~~$\nexists~~ g^{-1}: \mathcal{W} \to \mathcal{R}$, where $\mathcal{W}$ is the set of all observed watch time values, $\mathcal{G}$ is the function space, $\mathcal{R}$ is the set of all interest probability values.

% The correction function for estimating user interest is $r = f(w_{u,v})$. $\forall f\in \mathcal{F}$ defined in $w_{u,v}$ cannot allow $\forall w$ to find a counterpart $r_{u,v}$.
\end{theorem}

The proof of this theorem is presented in our Appendix~\ref{appendix: proof thm 1}. This theorem indicates that existing methods fail to uncover user interest in those completely played records, especially when the video duration is short. The failure of current methods motivates us to develop a CWT-based approach to address this problem and better understand user interest.

\section{Our approach}
\begin{figure}[t]
    \includegraphics[width=0.4\textwidth]{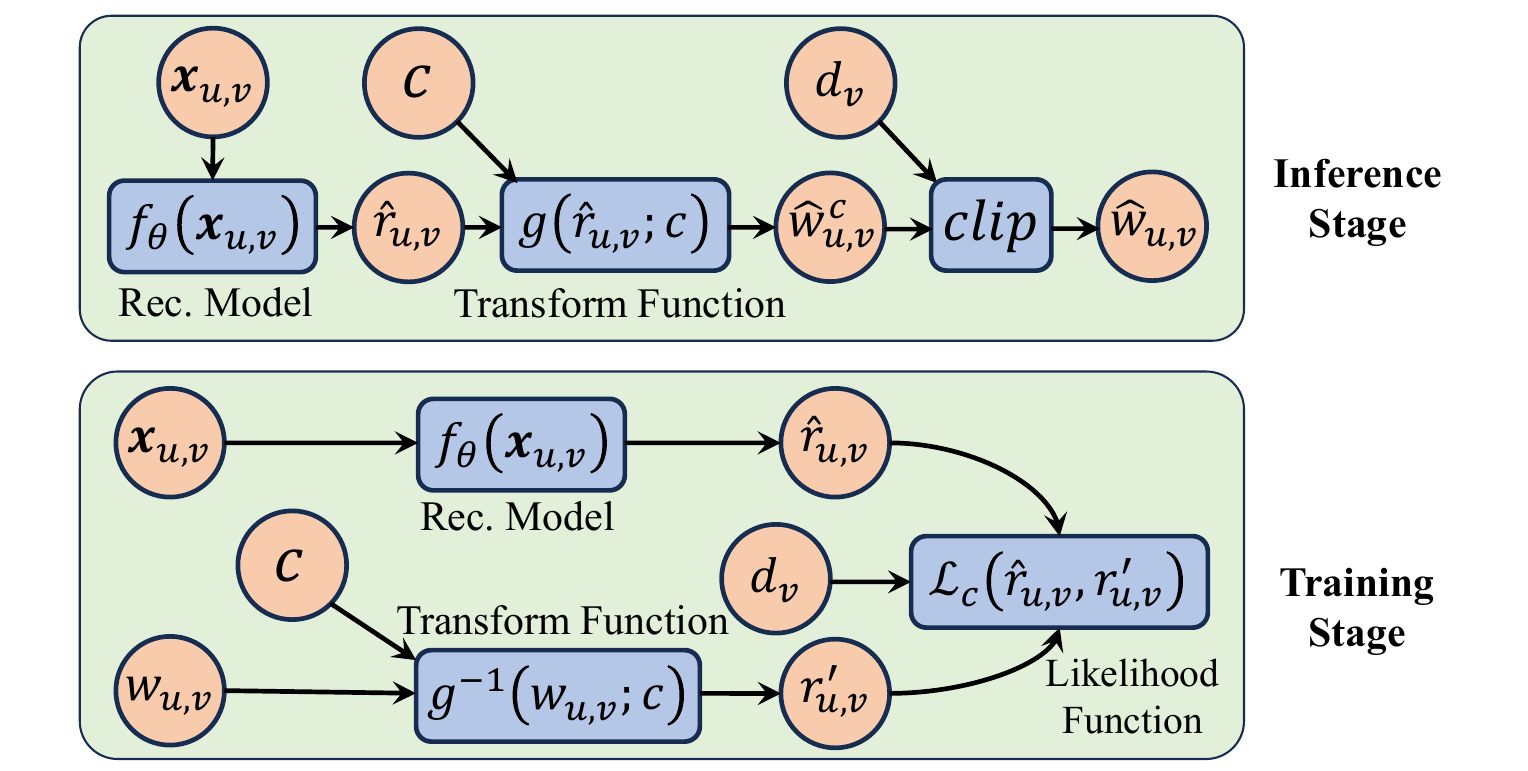}
    \caption{Computational flow of the proposed CWM.}
    \label{fig: alg pipeline}
\end{figure}

For better modeling the CWT and estimating user interest, we propose Counterfactual Watch Model (CWM). Fig~\ref{fig: alg pipeline} illustrates the flow of CWM in the inference stage and training stage. At the inference stage, a recommendation model $f_\theta(\cdot)$, parameterized with $\theta$, estimates the user interest $\hat{r}_{u,v}$ based on the feature vector $\mathbf{x}_{u,v}$. Then a transform function $g(\cdot)$, conditioned on user watch cost $c$, converts the interest estimation into the CWT prediction $\hat{w}^{c}_{u,v}$. The actual watch time prediction $\hat{w}_{u,v}$ is obtained by truncating it through the video duration $d_{v}$. 

At the training stage, to estimate the parameters $\theta$ in $f_\theta(\cdot)$, we employ a set of user activity log $\mathcal{D} \subseteq \{\mathcal{D}_{u,v}= \{\mathbf{x}_{u,v}, w_{u,v}, d_v\}: u\in \mathcal{U}, v\in\mathcal{V}\}$. For each $\mathcal{D}_{u,v} \in \mathcal{D}$, the observed watch time $w_{u,v}$ is transformed into the supervision signal of user interest by the inverse of the transform function, i.e., $ r^{\prime}_{u,v} =g^{-1}(w_{u,v}; c)$. Although we use observed watch time $w_{u,v}$ as the input of the inverse transform function here, we will approximate it to CWT in optimization. Then the predicted  user interest $\hat{r}_{u,v} = f_\theta(\mathbf{x}_{u,v})$ is calculated with current model parameters. For all $\mathcal{D}_{u,v} \in \mathcal{D}$, comparing $r^{\prime}_{u,v}$ and $\hat{r}_{u,v}$ jointly with duration $d_{v}$ derives the likelihood function $\mathcal{L}_c(\cdot)$. Thus, the learning of the recommendation model can be performed by optimizing $\mathcal{L}_c(\cdot)$. 

Next, we will elaborate on the design of CWM's key components: (1) the transform function $g(\cdot)$ and (2) the likelihood function $\mathcal{L}_c(\cdot)$.

%and its label $r^{\prime}_{u,v}$ is optimized by a likelihood function $\mathcal{L}_c(\cdot)$ joint with duration $d_{v}$. 

% We first develop a cost-based transform function which is defined to transform the counterfactual watch time into the estimation of user interest. Then the recommendation model can be learned by optimizing a counterfactual likelihood function defined over observed user watch times. In the inference stage, our CWM can not only estimate user interest for relevance ranking but also predict users' actual watch time more accurately.

\subsection{Cost-based transform function }
As defined in Section~\ref{sec: ct_wt define}, the CWT $w^{c}_{u,v}$ reveals user interest level $r_{u,v}$. To estimate $r_{u,v}$ from $w^{c}_{u,v}$, we first define the transform function between $w^{c}_{u,v}$ and $r_{u,v}$.

Based on the economic view of the user's watching process in Section~\ref{sec: economic view}, we first formulate the cumulative watch reward and cumulative watch cost function as:
\[
\begin{split}
    F_{\mathrm{reward}} &= \omega(r_{u,v}) \cdot \log(w^{c}_{u,v} + 1); \\
    F_{\mathrm{cost}} &= c \cdot w^{c}_{u,v},
\end{split}
\]
where the $\omega(r_{u,v})$ is the initial marginal reward that corresponds to user interest level $r_{u,v}$, and $c$ is the user watch cost per second. It is evident that the derivative function of $F_{\mathrm{reward}}$ is monotonically decreasing while that of $F_{\mathrm{cost}}$ is constant, which satisfies the assumptions in Section~\ref{sec: economic view}. When their derivative values are equal, we can derive a time point (i.e., CWT) for the maximum cumulative benefit:
\[
    \frac{\mathrm{d}~~ F_{\mathrm{reward}}}{\mathrm{d} ~~w^{c}_{u,v}} =  \frac{\mathrm{d} ~~F_{\mathrm{cost}}}{\mathrm{d} ~~w^{c}_{u,v}} \Rightarrow w^{c}_{u,v} = \frac{\omega(r_{u,v})}{c} - 1.
\]
Now we will formulate the initial marginal reward function $\omega(r_{u,v})$. Since $r_{u,v}\in(0,1)$, the $\omega(r_{u,v})$ is expected to fulfill the following conditions:(i) monotonically increasing.(ii) when $r_{u,v}\rightarrow 0$(i.e., the lowest interest level), the initial marginal reward $\omega(r_{u,v})$ should also tend to 0. (iii) when $r_{u,v}\rightarrow 1$(i.e., the highest interest level), the initial marginal reward $\omega(r_{u,v})$ should tend to positive infinity. To this end, we formulate that $\omega(r_{u,v}) = 1/(-\log r_{u,v})$, so the CWT can be further written as:
\begin{equation}
\label{eq: correct func}
    % w^{c}_{u,v} = g_{w^{c}}(r_{u,v};c) = \frac{1}{-c\log r_{u,v}} - 1.
    w^{c}_{u,v} = g(r_{u,v};c) = \frac{1}{-c\log r_{u,v}} - 1.
\end{equation}
Eq.~\eqref{eq: correct func} indicates how user interest $r_{u,v}$ and users' watch cost $c$ affect their CWT. Since we aim to uncover user interest from the CWT for training recommendation model, we can rewrite Eq.~\eqref{eq: correct func} to its inverse function:
\begin{equation}
\label{eq: correct func inverse}
    % r_{u,v} = g^{-1}(w^{c}_{u,v};c) = \exp\left(\frac{1}{-c (w^{c}_{u,v} + 1)}\right).
    r_{u,v} = g^{-1}(w^{c}_{u,v};c) = \exp\left(\frac{1}{-c (w^{c}_{u,v} + 1)}\right).
\end{equation}
Both Eq.~\eqref{eq: correct func} and Eq.~\eqref{eq: correct func inverse} are the cost-based transform functions since we introduce an extra cost parameter $c$ (as a hyper-parameter) for controlling the conversion sensitivity from CWT to user interest or vice versa. Then we can leverage them to estimate user interest and predict the user's actual watch time.
% That is, the higher the user watch cost corresponds to the higher user interest, in the case of equal counterfactual watch time.

\subsection{Counterfactual likelihood function}
Although we have proposed the cost-based transform function to describe the relationship between CWT and user interest, we still face the problem that CWT is truncated when the record is completely played. We need to approximate the CWT by the observed watch time to optimize our recommendation model.

\subsubsection{Formulation of the counterfactual likelihood function}
Inspired by the solution of survival analysis~\citep{Li2016Regularized}, we regard the observed watch time distribution as the truncated distribution of CWT. The overall likelihood function of the truncated CWT can be written as:
\begin{equation}
\label{eq: likelihood ct_watch}
\small
\begin{split}
    l &= \prod_{w^{c}_{u,v}<d_{v}} \Pr(W^{c} = w^{c}_{u,v}, W^{c}<d_{v}\mid \mathbf{x}_{u,v}) \prod_{w^{c}_{u,v}\geq d_{v}}  \Pr(W^{c}\geq d_{v}\mid \mathbf{x}_{u,v}) \\
    &= \prod_{w_{u,v}<d_{v}} \Pr(W^{c} = w_{u,v}, W^{c}<d_{v}\mid \mathbf{x}_{u,v})\prod_{w_{u,v}= d_{v}}  \Pr(W^{c}\geq d_{v}\mid \mathbf{x}_{u,v}),
\end{split}
\end{equation}
where $W^{c}$ denotes the random variable of CWT. Based on Eq.~\eqref{eq: cwt_and_awt}, we can replace all $w^{c}_{u,v}$ by $w_{u,v}$ in the second line. Eq~\eqref{eq: likelihood ct_watch} contains two parts: when $w_{u,v}<d_{v}$, the likelihood function equals the joint probability that the counterfactual duration variable is equal to $w_{u,v}$ and CWT variable is not truncated. When $w^{c}_{u,v}\geq d_{v}$, the likelihood function equals the probability that the CWT variable is truncated. As discussed in Theorem~\ref{thm: current limit}, we cannot find a transform function for indicating user interest from observed watch time. What we know is only $w^{c}_{u,v}\geq d_{v}$, so we incorporate this prior in likelihood function, allowing the models to determine the extent to which $w^{c}_{u,v}$ should exceed $d_{v}$.

Maximizing Eq.~\eqref{eq: likelihood ct_watch} can reduce the duration bias caused by the truncation of CWT. To estimate the user interest, we then equally transform Eq.~\eqref{eq: likelihood ct_watch} into the likelihood function of user interest $r_{u,v}$ via the transform function in Eq.~\eqref{eq: correct func inverse}:
\begin{small}
\begin{equation}
\label{eq: likelihood interest}
\begin{split}
    l &=\prod_{w_{u,v}<d_{v}} \Pr(R = g^{-1}(w_{u,v}; c),R < g^{-1}(d_{v}; c)\mid \mathbf{x}_{u,v}) \\
    &\times \prod_{w_{u,v}=d_{v}}\Pr(R\geq g^{-1}(d_{v};c)\mid \mathbf{x}_{u,v})\\
    & =\prod_{w_{u,v}<d_{v}} \Pr(R = g^{-1}(w_{u,v}; c)\mid \mathbf{x}_{u,v}) \prod_{w_{u,v}=d_{v}}\Pr(R\geq g^{-1}(d_{v};c)\mid \mathbf{x}_{u,v}),
\end{split}
\end{equation}
\end{small}
where $R$ denotes the random variable of user interest probability. Next, we will parameterize this likelihood function.

\subsubsection{Parameterize and optimize the likelihood function}
According to the result in ~\citep{Li2016Regularized}, Eq.~\eqref{eq: likelihood interest} can be parameterized with a theoretical guarantee if the random variable $R$ obeys Gaussian distribution. 
For converting $R$ into a Gaussian-distributed random variable, we employ an inverse function of the standard Gaussian-distributed cumulative distribution function:
% However, $R$ is a probability value and does not obey Gaussian distribution, we cannot directly parameterize above likelihood function. Hence, we first transform the interest estimation $g^{-1}(w_{u,v};c)$ and $g^{-1}(d_{v};c)$ into a Gaussian distributed random variable via the inverse function of standard Gaussian-distributed cumulative distribution function:

\[
% g^{\prime}(\cdot) = \Phi^{-1}\left[g^{-1}(\cdot)\right] \approx \log\left(g^{-1}(\cdot)/(1 - g^{-1}(\cdot))\right)
g^{\prime}(\cdot) = \Phi^{-1}\left[g^{-1}(\cdot)\right].
\]
Then we can parameterize Eq.~\eqref{eq: likelihood interest} into our counterfactual likelihood function:
\begin{equation}
\label{eq: likelihood parameterize}
\begin{split}
   \mathcal{L}_{c} = & \prod_{ w_{u,v}<d_{v}} \phi\left[\frac{g^{\prime}(w_{u,v};c) - f_\theta(\mathbf{x}_{u,v})}{\sigma}\right]\\
            & \times \prod_{w_{u,v}= d_{v}} \left(1 - \Phi\left[\frac{g^{\prime}(d_{v};c) - f_\theta(\mathbf{x}_{u,v})}{\sigma}\right]\right),
\end{split}
\end{equation}
%\begin{equation}
%\label{eq: likelihood parameterize}
%\begin{split}
    % l &= \prod_{w^{c}_{u,v}<d_{v}} \frac{\phi\left[\frac{g^{\prime}(w^{c}_{u,v}) - f_\theta(\mathbf{x}_{u,v})}{\sigma}\right]}{\Phi\left[\frac{g^{\prime}(d_{v}) - f_\theta(\mathbf{x}_{u,v})}{\sigma}\right]}\Phi\left[\frac{g^{\prime}(d_{v}) - f_\theta(\mathbf{x}_{u,v})}{\sigma}\right] \\
    % &*\prod_{w^{c}_{u,v}\geq d_{v}} \left(1 - \Phi\left[\frac{g^{\prime}(d_{v}) - f_\theta(\mathbf{x}_{u,v})}{\sigma}\right]\right) \\
    % &= \prod_{w^{c}_{u,v}<d_{v}} \phi\left[\frac{g^{\prime}(w^{c}_{u,v}) - f_\theta(\mathbf{x}_{u,v})}{\sigma}\right]  \prod_{w^{c}_{u,v}\geq d_{v}} \left(1 - \Phi\left[\frac{g^{\prime}(d_{v}) - f_\theta(\mathbf{x}_{u,v})}{\sigma}\right]\right) 
    %%%%%%%%%%%%%%%%%%%%%
    % \mathcal{L}_{c} = \prod_{w^{c}_{u,v}<d_{v}} \phi\left[\frac{g^{\prime}(w^{c}_{u,v}) - f_\theta(\mathbf{x}_{u,v})}{\sigma}\right]  \prod_{w^{c}_{u,v}\geq d_{v}} \left(1 - \Phi\left[\frac{g^{\prime}(d_{v}) - f_\theta(\mathbf{x}_{u,v})}{\sigma}\right]\right),
    %%%%%%%%%%%%%%%%%%%%%
%    \mathcal{L}_{c} = \left\{
%        \begin{aligned}
%            & \prod \phi\left[\frac{g^{\prime}(w_{u,v};c) - f_\theta(\mathbf{x}_{u,v})}{\sigma}\right],\quad \mathrm{if}~~ w_{u,v}<d_{v} \mathrm{;} \\
%            & \prod \left(1 - \Phi\left[\frac{g^{\prime}(d_{v};c) - f_\theta(\mathbf{x}_{u,v})}{\sigma}\right]\right), \quad \mathrm{if}~~ w_{u,v}= d_{v} \mathrm{;} 
%        \end{aligned}\right.
%\end{split}
%\end{equation}
where $\phi(\cdot)$ and $\Phi(\cdot)$\footnote{In practice, we approximate $\Phi(\cdot)$ via Sigmoid function follow~\citep{zhou2018jump}.} are the probability density function and cumulative distribution function of standard Gaussian distribution, respectively. $\sigma$ is the standard deviation of the interest, which is treated as a hyper-parameter in our method. And $f_\theta(\mathbf{x}_{u,v})$ is the recommendation model for predicting user interest, its parameter is denoted as $\theta$. Then the log-likelihood function is utilized in our training:
% The detailed derivation from Eq.\eqref{eq: likelihood interest} to Eq.~\eqref{eq: likelihood parameterize} can be found in our supplementary materials.
\begin{equation}
\label{eq: log likelihood parameterize}
\begin{split}
    \log(\mathcal{L}_{c}) = &\sum_{w_{u,v}<d_{v}} \frac{-\left( g^{\prime}(w_{u,v};c) - f_\theta(\mathbf{x}_{u,v})\right)^2}{2\sigma^2}\\
    & + \sum_{w_{u,v}=d_{v}} \log\Phi\left[\frac{f_\theta(\mathbf{x}_{u,v}) - g^{\prime}(d_{v};c)}{\sigma}\right].
\end{split}
\end{equation}

\textbf{Remark.} Eq.~\eqref{eq: log likelihood parameterize} is derived from maximum likelihood estimation (MLE): The optimal parameters are those that best describe the currently observed data. For the CWT, we observe that when the video is not fully watched, the CWT equals to the actual watch time, this is the MSE part of Eq.~\eqref{eq: log likelihood parameterize}. However, for the video is fully watched, the observation we only know is CWT may larger than the actual watch time, this led to the amplification part of Eq.~\eqref{eq: log likelihood parameterize}. This approach to modelling truncated data is also common in endogenous problems~\citep{amemiya1984tobit} and survival analysis~\citep{wang2019machine}. Although simply amplifies predictions for fully watched videos cannot precisely model interest, it can still enhance our interest predictions in a large margin, which is verified in our experimental results.

%\begin{equation}
%\label{eq: log likelihood parameterize}
%\begin{split}
%    \log(\mathcal{L}_{c}) = \left\{
%        \begin{aligned}
%            & \sum \frac{-\left( g^{\prime}(w_{u,v};c) - f_\theta(\mathbf{x}_{u,v})\right)^2}{2\sigma^2} \quad \mathrm{if}~~ w_{u,v}<d_{v} \mathrm{;} \\
%            & \sum \log\Phi\left[\frac{f_\theta(\mathbf{x}_{u,v}) - g^{\prime}(d_{v};c)}{\sigma}\right] ,\quad \mathrm{if}~~ w_{u,v}= d_{v} \mathrm{;} 
%        \end{aligned}\right.
%\end{split}
%\end{equation}
The detailed derivation from Eq.~\eqref{eq: likelihood interest} to Eq.~\eqref{eq: log likelihood parameterize} can be found in our appendix. Finally, a duration-debiased recommendation model can be obtained through maximizing Eq.~\eqref{eq: log likelihood parameterize}:
\[
    \theta^{*} \leftarrow \mathop{\arg\max}_{\theta}~\log(\mathcal{L}_{c}).
\]
% In this paper, we adopt Adam optimizer for model training.

\subsection{Online inference}
% The overall computational pipeline of our proposed CWM is presented in Fig~\ref{fig: alg pipeline}. 
In the inference stage, given a user-video pair $(u,v)$, the predicted interest and watch time can be calculated by the unbiased recommendation model $f_{\theta^*}$ parameterized by the learned parameter $\theta^*$:
\[
\begin{split}
    \hat{r}_{u,v} &= \Phi\left(f_{\theta^{*}}(\mathbf{x}_{u,v})\right); \\
    \hat{w}_{u,v} &= \mathrm{clip}\left(\frac{1}{-c\log \hat{r}_{u,v}} - 1,0,d_{v}\right),
\end{split}
\]
where $clip(x,a,b)$ function means clipping the value of $x$ into $[a,b]$.

% \begin{equation}
% \label{eq: likelihood parameterize}
% \begin{split}
%     l &= \prod_{w^{c}_{u,v}<d_{v}} \frac{\phi\left[(g^{\prime}(w^{c}_{u,v}) - f_\theta(\mathbf{x}_{u,v}))/\sigma\right]}{\Phi\left[(g^{\prime}(d_{v}) - f_\theta(\mathbf{x}_{u,v}))/\sigma\right]}\Phi\left[(g^{\prime}(d_{v}) - f_\theta(\mathbf{x}_{u,v}))/\sigma\right] \\
%     &*\prod_{w^{c}_{u,v}\geq d_{v}} \left(1 - \Phi\left[(g^{\prime}(d_{v}) - f_\theta(\mathbf{x}_{u,v}))/\sigma\right]\right) \\
%     &= \prod_{w^{c}_{u,v}<d_{v}} \phi\left[(g^{\prime}(w^{c}_{u,v}) - f_\theta(\mathbf{x}_{u,v}))/\sigma\right]  \prod_{w^{c}_{u,v}\geq d_{v}} \left(1 - \Phi\left[(g^{\prime}(d_{v}) - f_\theta(\mathbf{x}_{u,v}))/\sigma\right]\right) 
% \end{split}
% \end{equation}

% Without loss of generality, we assume the random variable $W^{c}$ obeys a Gaussian distribution, then Eq.~\eqref{eq: likelihood ct_watch} can be further parameterized as:
% \begin{equation}
% \label{eq: likelihood ct_watch}
% \begin{split}
%     l &= \prod_{w^{c}_{u,v}<d_{v}} \frac{\phi\left[(w_{u,v} - f(\mathbf{x}_{u,v}))/\sigma\right]}{\Phi\left[(d_{v} - f(\mathbf{x}_{u,v}))/\sigma\right]}\Phi\left[(d_{v} - f(\mathbf{x}_{u,v}))/\sigma\right] \\
%     &*\prod_{w^{c}_{u,v}\geq d_{v}} \left(1 - \Phi\left[(d_{v} - f(\mathbf{x}_{u,v}))/\sigma\right]\right) \\
%     &= \prod_{w^{c}_{u,v}<d_{v}} \phi\left[(w_{u,v} - f(\mathbf{x}_{u,v}))/\sigma\right]  \prod_{w^{c}_{u,v}\geq d_{v}} \left(1 - \Phi\left[(d_{v} - f(\mathbf{x}_{u,v}))/\sigma\right]\right) 
% \end{split}
% \end{equation}

\section{Experiments and Results}
\subsection{Experimental setting}
We conducted experiments to verify the effectiveness of CWM on two large-scale publicly available benchmarks and a dataset collected from an industrial video product. More implementation details can be found in Appendix~\ref{appendix: experimental setting}. More experimental results can be found in Appendix~\ref{appendix: more experimental results}. The source code and dataset are available at \textcolor{blue}{\url{https://github.com/hyz20/CWM.git}}.
\begin{table}[t]
\caption{Statistics of the datasets adopted in this study}
\resizebox{0.45\textwidth}{!}{
\begin{tabular}{cccccc}
\hline
Dataset  & \#Users & \#Videos & \#Interactions  & Mean Complete Ratio(\%)\\
\hline\hline
KuaiRand & 26,988  & 6,598    & 1,266,560       & 17.5\%   \\
WeChat   & 20,000  & 96,418   & 7,310,108       &  45.5\%   \\
Product   & 2,000,000  & 1,011,007  & 36,366,437        & 32.8\%   \\ \hline
% Dataset  & \#Users & \#Videos & \#Interactions & Duration Ranges(s) \\ \hline
% KuaiRand & 26,988  & 6,598    & 1,266,560      & {[}5,240{]}     \\
% WeChat   & 20,000  & 96,418   & 7,310,108      & {[}5,60{]}     \\
% Product   & 2,000,000  & 1,011,007  & 36,366,437      & {[}8,300{]}   \\ \hline
\end{tabular}}
\label{tab: dataset statistics}
\end{table}

\begin{table*}[t]
\caption{The watch time prediction performance of CWM and other baselines in KuaiRand, WeChat and Product. \textbf{Boldface} means the best-performed methods, while {\underline{underline}} means the second best-performed methods, superscripts $\dag$ means the significance compared to the second best-performed methods with $p<0.05$ of one-tailed $t$-test. `$\downarrow$' denotes that lower is better for MAE, while higher is better for XAUC.}
\resizebox{1.0\textwidth}{!}{
\begin{tabular}{c|c|ccccccc|ccccccc|ccccccc}
\hline
\multirow{2}{*}{Dataset}  & Backbone        & \multicolumn{7}{c|}{FM}                                                                                           & \multicolumn{7}{c|}{DCN}                                                                                       & \multicolumn{7}{c}{AutoInt}                                                                                            \\ \cline{3-23} 
                          & Method          & VR          & PCR         & WLR    & D2Q          & WTG    & \multicolumn{1}{c|}{D2Co}   & CWM                    & VR     & PCR         & WLR    & D2Q            & WTG    & \multicolumn{1}{c|}{D2Co}   & CWM                    & VR          & PCR         & WLR    & D2Q          & WTG         & \multicolumn{1}{c|}{D2Co}   & CWM                    \\ \hline
\multirow{2}{*}{KuaiRand} & MAE$\downarrow$ & 22.100      & 20.974      & 24.279 & {\ul 18.271} & 23.044 & \multicolumn{1}{c|}{22.262} & \textbf{17.738$^\dag$} & 21.698 & 21.910      & 21.044 & {\ul 18.131}   & 21.318 & \multicolumn{1}{c|}{21.564} & \textbf{17.420$^\dag$} & 21.726      & 21.465      & 23.202 & {\ul 18.213} & 22.771      & \multicolumn{1}{c|}{21.826} & \textbf{17.462$^\dag$} \\
                          & XAUC            & 0.683       & {\ul 0.697} & 0.668  & 0.646        & 0.666  & \multicolumn{1}{c|}{0.662}  & \textbf{0.714$^\dag$}  & 0.692  & {\ul 0.696} & 0.680  & 0.649          & 0.695  & \multicolumn{1}{c|}{0.693}  & \textbf{0.715$^\dag$}  & 0.685       & {\ul 0.699} & 0.668  & 0.646        & 0.673       & \multicolumn{1}{c|}{0.667}  & \textbf{0.713$^\dag$}  \\ \hline
\multirow{2}{*}{WeChat}   & MAE$\downarrow$ & 9.404       & 8.920       & 9.653  & {\ul 8.778}  & 9.637  & \multicolumn{1}{c|}{10.200} & \textbf{8.001$^\dag$}  & 9.317  & 8.709       & 9.149  & {\ul 7.987}    & 8.727  & \multicolumn{1}{c|}{8.667}  & \textbf{7.902}         & 9.405       & 8.910       & 9.815  & {\ul 8.815}  & 9.623       & \multicolumn{1}{c|}{10.031} & \textbf{7.969$^\dag$}  \\
                          & XAUC            & {\ul 0.696} & 0.692       & 0.667  & 0.682        & 0.653  & \multicolumn{1}{c|}{0.627}  & \textbf{0.713$^\dag$}  & 0.703  & 0.717       & 0.684  & \textbf{0.721} & 0.718  & \multicolumn{1}{c|}{0.718}  & 0.717                  & {\ul 0.700} & 0.693       & 0.661  & 0.679        & 0.653       & \multicolumn{1}{c|}{0.634}  & \textbf{0.714$^\dag$}  \\ \hline
\multirow{2}{*}{Product}  & MAE$\downarrow$ & 9.411       & 7.395       & 8.913  & {\ul 7.383}  & 7.420  & \multicolumn{1}{c|}{7.511}  & \textbf{6.785$^\dag$}  & 9.384  & {\ul 7.320} & 8.887  & 7.374          & 7.398  & \multicolumn{1}{c|}{7.443}  & \textbf{6.648$^\dag$}  & 9.192       & {\ul 7.073} & 8.668  & 7.169        & 7.125       & \multicolumn{1}{c|}{7.293}  & \textbf{6.591$^\dag$}  \\
                          & XAUC            & 0.808       & 0.801       & 0.816  & {\ul 0.817}  & 0.816  & \multicolumn{1}{c|}{0.812}  & \textbf{0.833$^\dag$}  & 0.808  & 0.803       & 0.817  & {\ul 0.823}    & 0.820  & \multicolumn{1}{c|}{0.817}  & \textbf{0.841$^\dag$}  & 0.810       & 0.801       & 0.810  & 0.819        & {\ul 0.820} & \multicolumn{1}{c|}{0.818}  & \textbf{0.835$^\dag$}  \\ \hline
\end{tabular}}
\label{tab: reg_result}
\end{table*}

\begin{table*}[]
\caption{The relevance ranking performance of CWM and other baselines in KuaiRand, WeChat and Product.}
\resizebox{1.0\textwidth}{!}{
\begin{tabular}{c|c|cccccccc|cccccccc|cccccccc}
\hline
\multirow{2}{*}{Dataset}  & Backbone & \multicolumn{8}{c|}{FM}                                                                                                                          & \multicolumn{8}{c|}{DCN}                                                                                                                               & \multicolumn{8}{c}{AutoInt}                                                                                                                \\ \cline{3-26} 
                          & Method   & \multicolumn{1}{c|}{Oracle} & VR          & PCR         & D2Q   & WTG   & NDT         & \multicolumn{1}{c|}{D2Co}        & CWM                   & \multicolumn{1}{c|}{Oracle} & VR          & PCR         & D2Q   & WTG         & NDT         & \multicolumn{1}{c|}{D2Co}        & CWM                   & \multicolumn{1}{c|}{Oracle} & VR    & PCR         & D2Q   & WTG         & NDT         & \multicolumn{1}{c|}{D2Co}  & CWM                   \\ \hline
\multirow{2}{*}{KuaiRand} & AUC      & \multicolumn{1}{c|}{0.738}  & 0.661       & 0.686       & 0.679 & 0.684 & 0.668       & \multicolumn{1}{c|}{{\ul 0.688}} & \textbf{0.735$^\dag$} & \multicolumn{1}{c|}{0.745}  & 0.665       & 0.693       & 0.684 & 0.718       & 0.679       & \multicolumn{1}{c|}{{\ul 0.730}} & \textbf{0.735}        & \multicolumn{1}{c|}{0.737}  & 0.655 & 0.691       & 0.679 & {\ul 0.692} & 0.669       & \multicolumn{1}{c|}{0.688} & \textbf{0.734$^\dag$} \\
                          & nDCG@3   & \multicolumn{1}{c|}{0.489}  & 0.442       & {\ul 0.469} & 0.464 & 0.462 & 0.461       & \multicolumn{1}{c|}{0.464}       & \textbf{0.486$^\dag$} & \multicolumn{1}{c|}{0.497}  & 0.448       & 0.469       & 0.466 & 0.471       & 0.464       & \multicolumn{1}{c|}{{\ul 0.480}} & \textbf{0.484}        & \multicolumn{1}{c|}{0.490}  & 0.444 & {\ul 0.469} & 0.466 & 0.466       & 0.459       & \multicolumn{1}{c|}{0.464} & \textbf{0.484$^\dag$} \\ \hline
\multirow{2}{*}{WeChat}   & AUC      & \multicolumn{1}{c|}{0.711}  & 0.639       & {\ul 0.651} & 0.645 & 0.602 & 0.629       & \multicolumn{1}{c|}{0.573}       & \textbf{0.703$^\dag$} & \multicolumn{1}{c|}{0.712}  & 0.641       & 0.699       & 0.696 & {\ul 0.703} & 0.672       & \multicolumn{1}{c|}{0.702}       & \textbf{0.707}        & \multicolumn{1}{c|}{0.714}  & 0.637 & {\ul 0.652} & 0.642 & 0.603       & 0.636       & \multicolumn{1}{c|}{0.575} & \textbf{0.704$^\dag$} \\
                          & nDCG@3   & \multicolumn{1}{c|}{0.588}  & 0.520       & {\ul 0.540} & 0.540 & 0.528 & 0.510       & \multicolumn{1}{c|}{0.527}       & \textbf{0.581$^\dag$} & \multicolumn{1}{c|}{0.589}  & 0.517       & {\ul 0.581} & 0.577 & 0.576       & 0.527       & \multicolumn{1}{c|}{0.570}       & \textbf{0.584}        & \multicolumn{1}{c|}{0.590}  & 0.519 & {\ul 0.542} & 0.540 & 0.534       & 0.512       & \multicolumn{1}{c|}{0.532} & \textbf{0.583$^\dag$} \\ \hline
\multirow{2}{*}{Product}  & AUC      & \multicolumn{1}{c|}{0.669}  & 0.605       & 0.593       & 0.623 & 0.624 & {\ul 0.640} & \multicolumn{1}{c|}{0.629}       & \textbf{0.660$^\dag$} & \multicolumn{1}{c|}{0.678}  & 0.608       & 0.593       & 0.625 & 0.626       & {\ul 0.649} & \multicolumn{1}{c|}{0.636}       & \textbf{0.663$^\dag$} & \multicolumn{1}{c|}{0.672}  & 0.610 & 0.593       & 0.629 & 0.631       & {\ul 0.644} & \multicolumn{1}{c|}{0.634} & \textbf{0.663$^\dag$} \\
                          & nDCG@3   & \multicolumn{1}{c|}{0.589}  & {\ul 0.556} & 0.472       & 0.513 & 0.523 & 0.553       & \multicolumn{1}{c|}{0.531}       & \textbf{0.582$^\dag$} & \multicolumn{1}{c|}{0.595}  & {\ul 0.564} & 0.475       & 0.516 & 0.533       & 0.563       & \multicolumn{1}{c|}{0.540}       & \textbf{0.591$^\dag$} & \multicolumn{1}{c|}{0.591}  & 0.558 & 0.479       & 0.514 & 0.527       & {\ul 0.559} & \multicolumn{1}{c|}{0.533} & \textbf{0.585$^\dag$} \\ \hline
\end{tabular}}
\label{tab: rank_result}
\end{table*}

\subsubsection{Datasets}
The experiments were conducted on two public real-world datasets: WeChat and KuaiRand. They are respectively collected from two large micro-video platforms, Wechat Channels and Kuaishou. We also conduct our evaluation in a large-scale product dataset from our video platform, which has tens of billions of daily active users. We list their statistic information in Table~\ref{tab: dataset statistics}. Note that we present each dataset's completely played record percentage in the last column of Table~\ref{tab: dataset statistics}. Since the duration bias usually occurs on those completely played records, their percentages in all records represent the severity of the duration bias of each dataset. According to the statistics, WeChat has the most serious bias, while KuaiRand has the least bias. 

% \textbf{WeChat}. This dataset is released by WeChat Big Data Challenge 2021, containing the Wechat Channels logs within two weeks. Following the practice in \citep{Zheng2022DVR}, we split the data into the first 10 days, the middle 2 days, and the last 2 days as training, validation, and test set. The adopted input features include \emph{userid},\emph{feedid},\emph{device},\emph{authorid},\quad\emph{bgm\_song\_id},\emph{bgm\_singer\_id},\emph{user\_type}, \emph{like}, \emph{read\_comment}, \emph{forward}.

% \textbf{KuaiRand}~\citep{gao2022kuairand}. KuaiRand is a newly released sequential recommendation dataset collected from KuaiShou. As suggested in~\citep{gao2022kuairand}, we utilized one of the subsets \emph{KuaiRand-pure} in this study. To mitigate the sparsity problem, we selected data from which the video duration is up to 4 minutes. We split the data into the first 14 days, the middle 7 days, and the last 10 days as training, validation, and test set. The adopted input features include \emph{user\_id}, \emph{video\_id}, \emph{author\_id}, \emph{music\_id}, \emph{video\_type},\emph{upload\_type},\emph{tab},\emph{is\_like}, \emph{is\_follow}, \emph{is\_comment}, \emph{is\_forward}, \emph{is\_profile\_enter},\emph{is\_hate}, \emph{most\_popular\_tag}.

% $\frac{\mathrm{\# complete play}}{\# Interactions}$\%

% \begin{figure}[t]
% % \vspace{-3pt}
%     \includegraphics[width=0.25\textwidth]{figs/complete_rate.pdf}
%     \caption{completely played record percentage of each dataset.}
%     \label{fig: complete stat}
% \vspace{-6pt}
% \end{figure}

\subsubsection{Evaluation}
In this paper, we not only adopt our CWM for ranking videos by user interest (i.e., relevance ranking) but also for predicting users' watch time. Both tasks are of great importance in real video recommendation scenarios. As for the task of watch time prediction, we utilize users' actual watch time $w_{u,v}$ as the ground truth, \textbf{MAE}(Mean Absolute Error) and  \textbf{XAUC}~\citep{Zhan2022Deconfounding} were used as the evaluation measures. Note that XAUC evaluates if the predictions of two samples are in the same order as their actual watch time. Such pairs are uniformly sampled, and the percentile of samples that are correctly ordered by predictions is XAUC. A larger XAUC suggests better watch time prediction performance.
% and \textbf{XGAUC} the XGAUC is the metric that calculates XAUC per user and then averages scores with weights proportional to user sample size. Larger XGAUC suggests better model performance.

As for evaluating the task of relevance ranking according to user interest, considering that the user interest labels are unobserved in real-world datasets, we defined user interests based on CWT. Given a $(u,v)$ pair, the user interest label is defined as:
% we resort to the definition of \emph{long\_view} from the KuaiRand dataset~\citep{gao2022kuairand} as the user interest indicator. Given a $(u,v)$ pair, the user interest label is defined as:
\begin{equation}    \label{eq: interest define}
\begin{small}
r_{u,v} = 
    \begin{cases}
    1 & \text{if }(d_{v}\leq w_{\text{0.7}} \land w_{u,v}\!=\!d_{v}) \!\lor (d_{v}\!>\!w_{\text{0.7}}\! \land\! w_{u,v}\!>\!w_{\text{0.7}}\!)\\
    0 & \text{otherwise.}\\
    \end{cases}
\end{small}
\end{equation}
The $w_{\text{0.7}}$ indicates the 70\% percentile of observed watch time, which is considered as the threshold for CWT. When a user watches beyond this time, we consider the user to be interested. The similar user interest definition is also adopted in~\citep{gao2022kuairand, Zhao2023Uncovering}. We will discuss the unbiasedness of $r_{u,v}$ in Appendix~\ref{appendix: The Unbiasedness of Interest Labels}. The $r_{u,v}$ is used as the ground truth for evaluating the relevance ranking task, \textbf{AUC} and \textbf{nDCG@k} are utilized as the evaluation metrics.

\subsubsection{Baselines}
In the experiments, we compared the proposed method with the following baselines: 
Three duration debiased baselines: \textbf{PCR}, \textbf{WTG}~\citep{Zheng2022DVR},\textbf{D2Q}~\citep{Zhan2022Deconfounding}, \textbf{D2Co}~\citep{Zhao2023Uncovering}
; Two watch time-weighted baselines: \textbf{WLR}~\citep{Covington2016Deep} and \textbf{NDT}~\citep{Xie2023Reweighting}; And a naive baseline: \textbf{VR} (value regression) which directly fit the observed watch time. For relevance ranking task, we also provide the result of \textbf{Oracle}, which is trained directly with the label defined in Eq.~\eqref{eq: interest define} and denote the upper bound performance of relevance ranking. Most of these methods are designed initially for relevance ranking or watch time prediction via a transform function. For the relevance ranking task, we directly rank the candidate videos via the predicted scores output by the recommendation models trained by these methods. For the watch time prediction task, we first transform the prediction of recommendation models into the interval of watch time via the inverse transform function of each method. Then we clip the estimated watch time into 0 to $d_{v}$ as what we did in CWM. Two exceptions are WLR and NDT, which are implemented by a two-tower model and have no inverse transform function available. Therefore, WLR is only used in watch time prediction, and NDT is used only in relevance ranking. 

To investigate the generalization of our method and the baselines, we integrate them with different backbone models. Specifically, we use recommendation models of \textbf{FM}~\citep{Rendle2012Factorization}, \textbf{DCN}~\citep{Wang2017Deep} and \textbf{AutoInt}~\citep{Song2019AutoInt} as the backbone models. These three backbone models respectively represent three types of feature interactions: inner product, outer product, and attention mechanisms. 

% \subsubsection{Implementation Details}
% We implement all the backbones with pytorch-fm\footnote{\url{https://github.com/rixwew/pytorch-fm}}, an open-source library for factorization machine models. In particular, we implement WLR following the details in ~\citep{Zhan2022Deconfounding}. For D2Q, the group number is set to 60 on KuaiRand and 30 on WeChat and 18 in our Product dataset. For NDT, we set its hyper-parameters as author suggested in their paper. We utilize Adam as the optimizer and set the initial learning rate as $5e-4$ for all methods. The batch size is set as 512. For all the backbone models, we set their latent embedding dimension to 10. For all methods with neural networks, the hidden units are set to 64 while the dropout ratio is set to 0.2. The value of user cost $c$ in our CWM is set to ${0.025,0.025, 0.2}$, and the value of $\sigma$ is set to $\{2, 20, 5\}$ on KuaiRand, WeChat and Product dataset, respectively. We tune our hyper parameters on the validation set while evaluating the performance on the test set. The source code is available at \url{https://xxxx/xxxx/xxxx}.

\subsection{Overall performance}
We compared our CWM with other baselines in the three datasets' watch time prediction task and relevance ranking task, as shown in Table~\ref{tab: reg_result} and Table~\ref{tab: rank_result}, respectively. It can be seen that our CWM obtains the best performance on almost all three datasets, all backbones and both tasks significantly. We also note that on WeChat, those methods equipped with duration debiasing (e.g., WTG and D2Q) perform even worse than the naive VR method. The reason is that the WeChat dataset has much more completely played records (45.5\%) than that of KuaiRand (17.5\%), showing that increasing completely played records make current debiasing methods get ineffective. In contrast, CWM improved more on WeChat than on other datasets. The results also verified the motivation of this paper: Current methods regard all completely played records as the same high interest, violating real interest distribution in real data. Therefore, when the dataset contains many completely played records (i.e., records with truncated CWT), the performance of these methods gets worse. Instead, CWM can model users' truncated CWT to estimate user interest better and predict users' actual watch time.

\begin{figure}
    \subfigure[watch time prediction]{
    \includegraphics[width=0.22\textwidth]{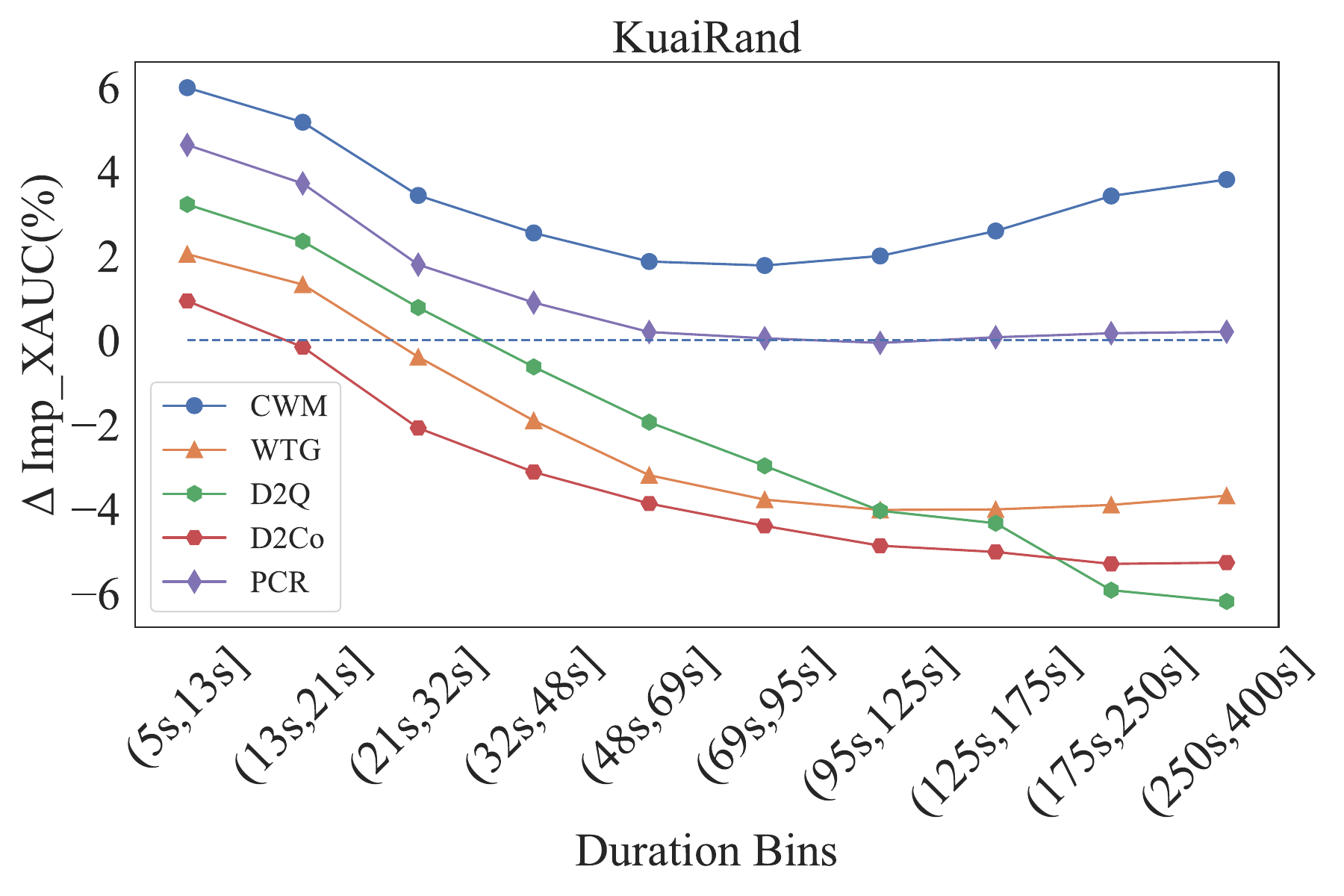}}
    \quad
    \subfigure[relevance ranking]{
    \includegraphics[width=0.22\textwidth]{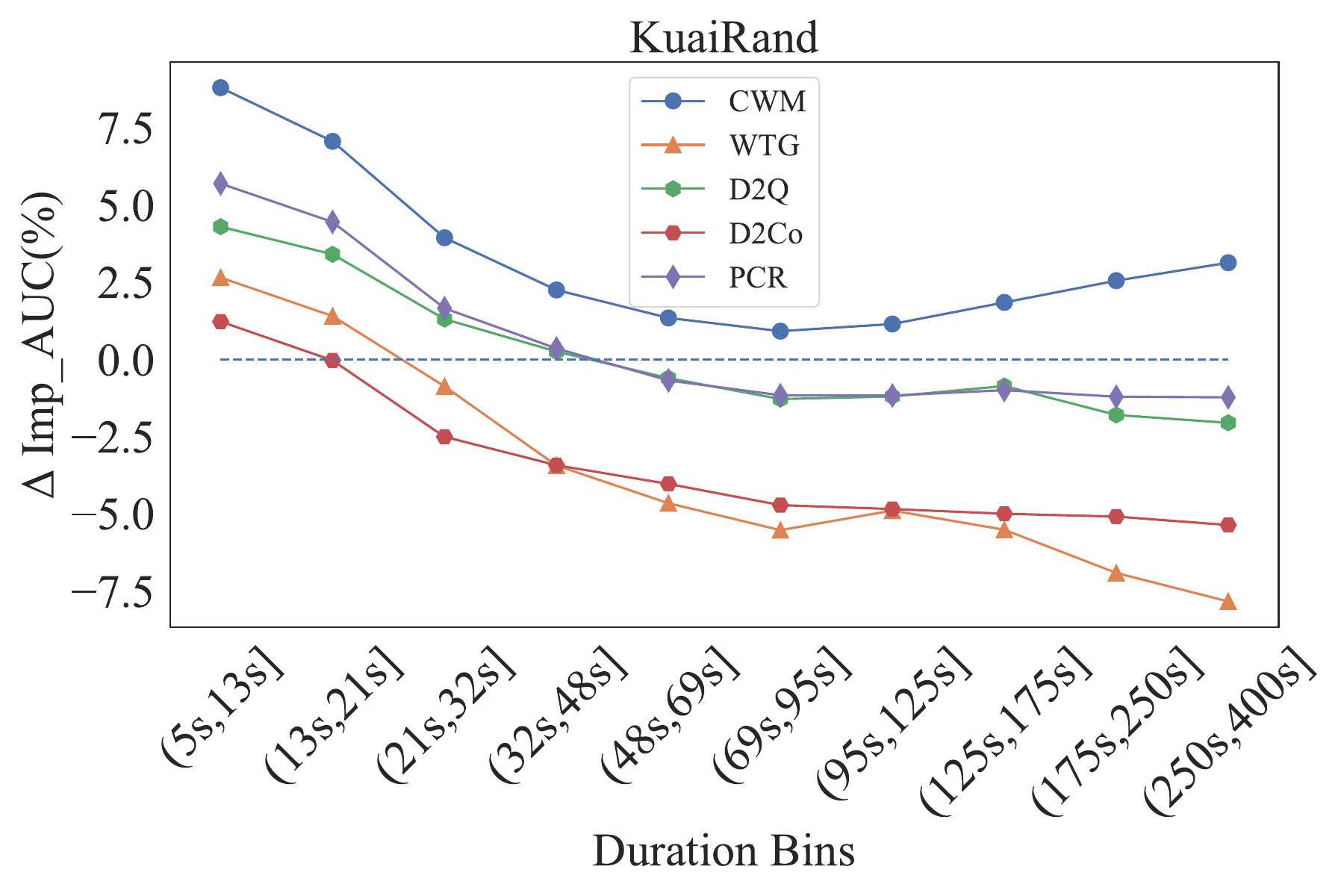}}
    \caption{The relative improvement of each methods to VR on different duration bins of KuaiRand.}
    % (a) relative improvement on XAUC. (b) relative improvement on AUC. The backbone model is FM.
    \label{fig: group analysis}
\end{figure}

% \begin{figure}[t]
%     \includegraphics[width=0.45\textwidth]{figs/CWM_fenduan.pdf}
%     \caption{The relative improvement of each methods to VR on different duration bins of KuaiRand and WeChat. Left two: watch time prediction task. Right two: relevance ranking task.}
%     \label{fig: group analysis}
% \end{figure}

\subsection{Effectiveness on duration debiasing}
To investigate why our CWM is more effective on duration debiasing than other baselines, we divided the KuaiRand dataset into ten equal parts with different duration ranges. Then we evaluate each model on the subset of KuaiRand. The result is presented in Fig.~\ref{fig: group analysis}. Note that the evaluation metric has different value scales among different subsets, so we report the relative improvement of each method to VR to show the extent to which these methods address duration bias. The relative improvement is $\Delta Imp = \frac{v_m - v_0}{v_0}$, where $v_0$ is the metric value of VR and $v_m$ is the metric value of each method. 

Fig~\ref{fig: group analysis}(a) illustrates the performance of each method on the watch time prediction task, measured by $\Delta Imp $ on XAUC. Fig~\ref{fig: group analysis}(b) illustrates the performance of each method on the relevance ranking task, measured by $\Delta Imp $ on AUC. In both tasks, most baselines perform better than VR in short videos (i.e., duration<30s), indicating their effectiveness on duration debiasing to some extent. However, since they simply regard completely played records as equally high interest, CWM performed better than them. We can find that these baselines perform worse in longer videos on both tasks. This is also because they regard short, completely played video recordings as high interest, leading to underestimating user interest and watch time prediction for longer videos. In contrast, CWM can model the CWT and assign fairer interest estimates to videos of different durations.

\begin{figure}[t]
    \includegraphics[width=0.45\textwidth]{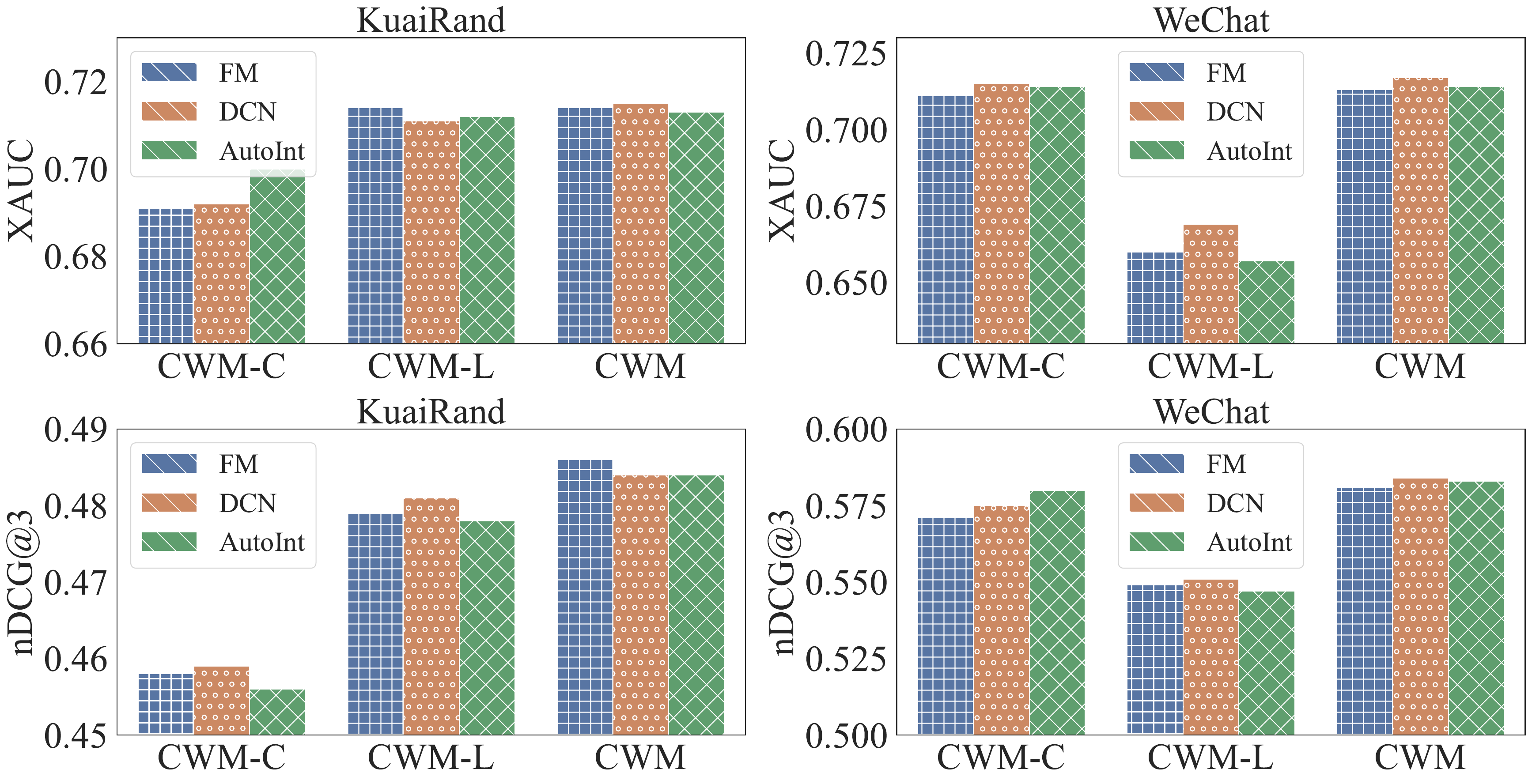}
    \caption{Ablation study on CWM in both KuaiRand and WeChat. Top two: watch time prediction task. Bottom two: relevance ranking task.}
    \label{fig: ablation study}
\end{figure}

\subsection{Comparison with more baselines}
We have compared our CWM with other label-correction methods before, but the superiority of CWM compared with other state-of-the-art methods remains unclear. Therefore, we further implemented CVRDD~\citep{Tang2023Video}, DCR~\citep{He2023Confounding}, and VLDRec~\citep{Quan2023Video}. CVRDD treats video duration as a mediation factor, DCR considers video duration as a confounder, and VLDRec employs PCR as the label with pairwise learning, all primarily feature-based or data-based approaches. In contrast, our label-correction method, CWM, consistently outperforms these baselines in terms of AUC and nDCG@3 metrics, as shown in Table~\ref{tab: more_result}.
% For instance, in the KuaiRand dataset with the FM backbone, CWM achieves an AUC of 0.735 and an nDCG@3 of 0.486, surpassing the best baseline, DCR, which records 0.695 and 0.469, respectively. Similarly, in the WeChat dataset, CWM with the DCN backbone scores an AUC of 0.707 and an nDCG@3 of 0.584, outperforming DCR's 0.681 and 0.559. 
The superiority of CWM is consistent across all datasets and backbone models. In conclusion, our CWM method demonstrates superior performance in mitigating duration bias by modeling counterfactual watch time (CWT), highlighting its effectiveness in delivering more accurate and unbiased recommendations in short video recommendation systems.
\begin{table}[]
\caption{The ranking performance of CWM and more state-of-the-art baselines in KuaiRand, WeChat and Product.}
\resizebox{0.48\textwidth}{!}{
\begin{tabular}{cc|cc|cc|cc}
\hline
\multicolumn{2}{c|}{Dataset}                           & \multicolumn{2}{c|}{KuaiRand}   & \multicolumn{2}{c|}{WeChat}     & \multicolumn{2}{c}{Product}     \\ \hline
\multicolumn{1}{c|}{Backbone}                 & Method & AUC            & nDCG@3         & AUC            & nDCG@3         & AUC            & nDCG@3         \\ \hline
\multicolumn{1}{c|}{\multirow{5}{*}{FM}}      & Oracle & 0.738          & 0.489          & 0.711          & 0.588          & 0.669          & 0.589          \\ \cline{2-8} 
\multicolumn{1}{c|}{}                         & VLDRec & 0.689          & 0.463          & 0.655          & 0.540          & 0.633          & {\ul 0.543}    \\
\multicolumn{1}{c|}{}                         & DCR    & {\ul 0.695}    & {\ul 0.469}    & 0.659          & 0.542          & {\ul 0.639}    & 0.540          \\
\multicolumn{1}{c|}{}                         & CVRDD  & 0.682          & 0.462          & {\ul 0.662}    & {\ul 0.545}    & 0.629          & 0.535          \\ \cline{2-8} 
\multicolumn{1}{c|}{}                         & CWM    & \textbf{0.735$^\dag$} & \textbf{0.486$^\dag$} & \textbf{0.703$^\dag$} & \textbf{0.581$^\dag$} & \textbf{0.660$^\dag$} & \textbf{0.582$^\dag$} \\ \hline
\multicolumn{1}{c|}{\multirow{5}{*}{DCN}}     & Oracle & 0.745          & 0.497          & 0.712          & 0.589          & 0.678          & 0.595          \\ \cline{2-8} 
\multicolumn{1}{c|}{}                         & VLDRec & 0.701          & 0.468          & 0.679          & 0.547          & 0.647          & 0.543          \\
\multicolumn{1}{c|}{}                         & DCR    & {\ul 0.716}    & {\ul 0.480}    & 0.681          & {\ul 0.559}    & {\ul 0.649}    & {\ul 0.554}    \\
\multicolumn{1}{c|}{}                         & CVRDD  & 0.684          & 0.473          & {\ul 0.687}    & 0.556          & 0.630          & 0.535          \\ \cline{2-8} 
\multicolumn{1}{c|}{}                         & CWM    & \textbf{0.735$^\dag$} & \textbf{0.484} & \textbf{0.707$^\dag$} & \textbf{0.584$^\dag$} & \textbf{0.663$^\dag$} & \textbf{0.591$^\dag$} \\ \hline
\multicolumn{1}{c|}{\multirow{5}{*}{AutoInt}} & Oracle & 0.737          & 0.490          & 0.714          & 0.590          & 0.672          & 0.591          \\ \cline{2-8} 
\multicolumn{1}{c|}{}                         & VLDRec & 0.698          & 0.472          & 0.661          & 0.546          & 0.638          & {\ul 0.550}    \\
\multicolumn{1}{c|}{}                         & DCR    & {\ul 0.699}    & {\ul 0.475}    & 0.662          & 0.547          & {\ul 0.643}    & 0.546          \\
\multicolumn{1}{c|}{}                         & CVRDD  & 0.683          & 0.470          & {\ul 0.664}    & {\ul 0.555}    & 0.639          & 0.536          \\ \cline{2-8} 
\multicolumn{1}{c|}{}                         & CWM    & \textbf{0.734$^\dag$} & \textbf{0.484} & \textbf{0.704$^\dag$} & \textbf{0.583$^\dag$} & \textbf{0.663$^\dag$} & \textbf{0.585$^\dag$} \\ \hline
\end{tabular}}
\label{tab: more_result}
\end{table}

\subsection{Ablation study}
We also investigate how CWM's two components benefit the CWM, i.e., the cost-based transform function and the counterfactual likelihood functions. The cost-based transform function estimates user interest from the CWT, and the counterfactual likelihood function optimizes the model unbiasedly using the observed watch time. Specifically, we produce two variants for CWM. The first is denoted as \textbf{CWM-C}, which removes the cost-based transform function and directly applies the original watch time to the counterfactual likelihood function. The second one is denoted as \textbf{CWM-L}, which replaces the counterfactual likelihood function with a mean squared error loss function. 

Fig.~\ref{fig: ablation study} demonstrates the performance comparison between CWM and its variants on two datasets and two tasks. On KuaiRand, CWM-L obtains similar performances to CWM, while CWM-C has a significant performance drop to CWM. We argue that when CWT is less truncated (e.g., KuaiRand has only 17.5\% completely played records), how it is converted into an interest estimation can primarily affect performance. On WeChat, CWM-C obtains a similar performance to CWM, while CWM-L has a significant performance drop to CWM. We argue that when CWT is heavily truncated (e.g., WeChat has 45.5\% completely played records), how the observed watch time is used to approximate the learning of CWT becomes the performance bottleneck.

\subsection{Online A/B Testing}
\label{sec: ab_test}
To verify the effectiveness of CWM in real-world recommendation scenarios, we conducted online experiments in our commercial system, a popular platform with tens of millions of active users every day. The baseline is a highly-optimized multi-task model deployed for the product.%, denoted as $\mathcal{M}_{base}$. 
Both the baseline and CWM were trained incrementally on the same anonymous logging data, and each one serves $5\%$ traffics, randomly selected from the same user group. As for short video recommendations, improving customers' mean watch time (MWT) is the main target. Other metrics, such as average valid viewing volume (VV) and click-through rate (CTR), are also adopted. % and play complete rate (PCR). %Among them, the valid VV is of more significance, which represents the number of videos whose corresponding dwell time exceeds a certain length. 
According to the online A/B testing results shown in Table~\ref{tab:online result}, we can see that CWM does help users to entertain themselves and spend more time watching the short videos.

\begin{table}[t]
\caption{Relative improvement of CWM to product baseline for one-week online A/B testing.}
\label{tab:online result}
\resizebox{0.28\textwidth}{!}{
\begin{tabular}{cc c cc}
    \hline
%    \multirow{2}{*}{Methods}&Main Metric.&&\multicolumn{3}{c}{Other Metrics.}\\
%    \cmidrule{2-2} \cmidrule{4-6}
     &MWT && VV & CTR \\
    \hline
    Product & - && - & -  \\
    CWM  &+2.9\%&&+2.5\%&+0.3\%\\
    \hline
\end{tabular}}
\end{table}

\section{Conclusion}
In this study, we aim to counteract the duration bias in video recommendation. We propose counterfactual watch time (CWT) for interpreting the duration bias in video recommendation and point out that the duration bias is caused by the truncation of the user's CWT by video duration. A Counterfactual Watch Model (CWM) is then developed, revealing that the CWT equals the time users get the maximum benefit from video recommender systems. A cost-based correction function is defined to transform the CWT into the user interest, and the unbiased recommendation model can be learned by optimizing a counterfactual likelihood function defined over observed user watch times. Experimental results on three offline real datasets and online A/B testing indicate the superiority of the proposed CWM.
% \input{Appendix.tex}

% \section*{Ethical Considerations}
% This paper proposes a counterfactual watch model for watch time prediction and relevance ranking, which are two important tasks in video recommendation. Our method follows the research route of existing works for enabling users to make better use of recommender systems. There are no negative societal impacts in our study.

\begin{acks}
This work was funded by the National Key R\&D Program of China (2023YFA1008704), the National Natural Science Foundation of China (No. 62377044), Beijing Key Laboratory of Big Data Management and Analysis Methods, Major Innovation \& Planning Interdisciplinary Platform for the “Double-First Class” Initiative, funds for building world-class universities (disciplines) of Renmin University of China, and PCC@RUC.
\end{acks}

\balance
\bibliographystyle{ACM-Reference-Format}
%\bibliography{sample-base}
%\bibliographystyle{plain}
\bibliography{ref}

\appendix

\section{Proof of theorem 1}
\label{appendix: proof thm 1}
\begin{proof}
    As we have analyzed before, the CWT $w^{c}_{u,v}$ corresponds to user interests $r_{u,v}$, i.e., $w^{c}_{u,v} = g(r_{u,v})$. Based on Eq.~\eqref{eq: cwt_and_awt}, we can obtain the functions of $w_{u,v}$ on $r_{u,v}$:
    \[
        w_{u,v} = \mathrm{min}(g(r_{u,v}), d_{v}),\quad r_{u,v}\in\mathcal{R},~~w_{u,v}\in\mathcal{W}
    \]
    Next, we need to prove that the above function does not always have an inverse function, w.r.t any $w_{u,v}$. Similar to Eq.~\eqref{eq: cwt_and_awt}, we rewrite the above function as a segmented function:
    \[
        \begin{aligned}
            &  r_{u,v}= g^{-1}(w_{u,v}),\quad \mathrm{if}~~ w_{u,v}<d_{v} \mathrm{;} \\
            & r_{u,v} \geq g^{-1}(w_{u,v}),\quad \mathrm{if}~~ w_{u,v}=d_{v} \mathrm{;} 
        \end{aligned}
    \]
    Note that when $w_{u,v}=d_{v}$ (i.e., completely play a video), we can only obtain an inequality between $w_{u,v}$ and $r_{u,v}$, thus proving that there is no such an inverse function $r_{u,v} = g^{-1}(w_{u,v})$ for all $w_{u,v}\in\mathcal{W}$. 
\end{proof}

\section{Detailed Experimental Setting}
\label{appendix: experimental setting}
In our paper, we briefly described the dataset used in our papers and the baselines due to the limitation of pages. Now, we put a more detailed experimental setting in this appendix.

% \subsection{Dataset description}
\textbf{WeChat}. This dataset is released by WeChat Big Data Challenge 2021, containing the WeChat Channels logs within two weeks. Following the practice in \citep{Zheng2022DVR}, we split the data into the first 10 days, the middle 2 days, and the last 2 days as training, validation, and test set. We noticed that there is an unusually high proportion of 60s videos among the dataset (17.3\%), so our experiments were conducted on a subset of 60s videos that were excluded (i.e., the duration range is [5s,59s] in the subset). The adopted input features include \emph{userid},\emph{feedid},\emph{device},\emph{authorid},\emph{bgm\_song\_id},\emph{bgm\_singer\_id},\emph{user\_type}.

\textbf{KuaiRand}~\citep{gao2022kuairand}. KuaiRand is a newly released sequential recommendation dataset collected from KuaiShou. As suggested in~\citep{gao2022kuairand}, we utilized one of the subsets \emph{KuaiRand-pure} in this study. To mitigate the sparsity problem, we selected data from which the video duration is up to 400s. We split the data into the first 14 days, the middle 7 days, and the last 10 days as training, validation, and test set. The adopted input features include \emph{user\_id}, \emph{video\_id}, \emph{author\_id}, \emph{music\_id},  \emph{follow\_user\_num\_range},\emph{register\_days\_range}, \emph{fans\_user\_num\_range}, \emph{friend\_user\_num\_range}, \emph{user\_active\_degree}, \emph{most\_popular\_tag},
\emph{video\_type},\emph{upload\_type},\emph{tab}.

\textbf{Product}. We collect the product dataset from the server log of our video platform, which samples the log data from June 19, 2023 to June 25, 2023. Due to the data imbalance, we intercepted the records below 42 seconds as the final training set. We trained our model on this dataset and tested them in the last two hours of server log data. The pretrained ID embedding and side information are used as feature inputs for all methods.

In this study, we implement WLR following the details in ~\citep{Zhan2022Deconfounding}. For D2Q, the group number is set to 60 in KuaiRand, 30 in WeChat, and 10 in our Product dataset. For NDT, we set its hyper-parameters as the author suggested in their paper~\citep{Xie2023Reweighting}. For D2Co, the window size and sensitivity-controlled term are set as suggested in their paper~\cite{Zhao2023Uncovering}. We utilize Adam as the optimizer and set the initial learning rate as $5e^{-4}$ for all methods. The batch size is set as 512. For all the backbone models, we set their latent embedding dimension to 10. For all methods with neural networks, the hidden units are set to 64 while the dropout ratio is set to 0.2. The value of user cost $c$ and $\sigma$ in our CWM is set to $(1/40,2)$ in the KuaiRand dataset, $(1/40,20)$ in the WeChat dataset, and $(1/5,5)$ in the Product dataset. We tune our hyperparameters on the validation set while evaluating the performance on the test set.

\begin{figure}
    \subfigure[KuaiRand]{
    \includegraphics[width=0.42\textwidth]{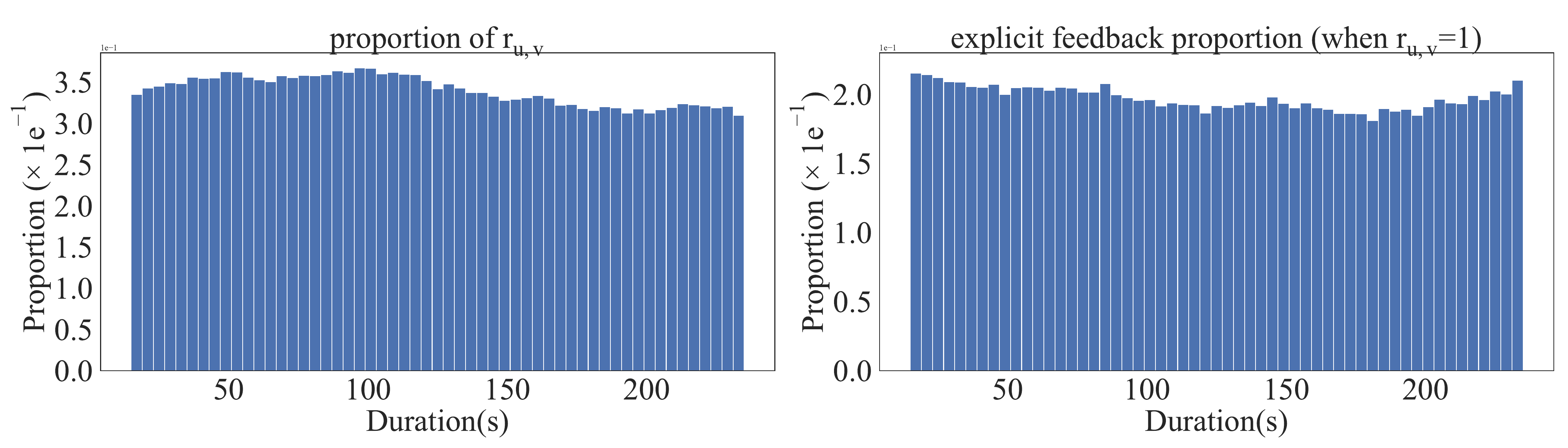}}
    \quad
    \subfigure[WeChat]{
    \includegraphics[width=0.42\textwidth]{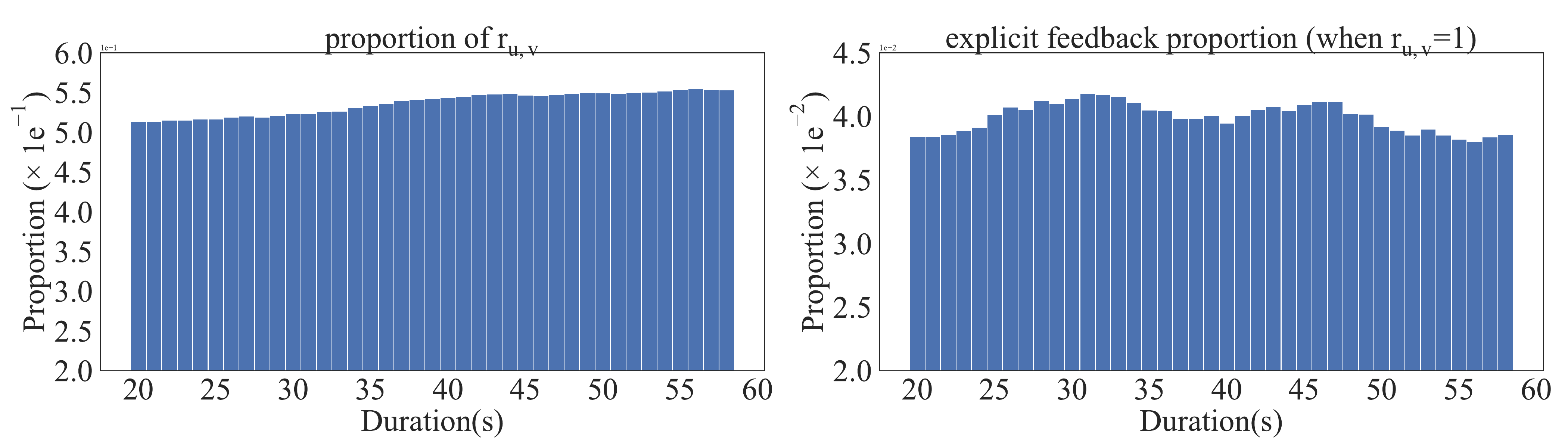}}
    \caption{(Left) The proportion of user interest labels $r_{u,v}$ and (Right) The proportion of explicit feedback when $r_{u,v}=1$ grouped by video duration on (a) KuaiRand and (b) WeChat.}
    \label{fig: long_view verify}
\end{figure}

\section{The Unbiasedness of Interest Labels}
\label{appendix: The Unbiasedness of Interest Labels}
To evaluate the performance of relevance ranking tasks in this study, we need an unbiased indicator of user interest first. However, the user interest labels are unobserved in real-world datasets. 
Although explicit feedback can reflect user interest and is not affected by duration bias, it suffers from severe selection bias and noise~\citep{Wang2022Denoising, Gao2022Denoising,Wang2021Denoising}: this indicates that users might not provide explicit feedback for videos they like and might mistakenly provide explicit feedback for videos they dislike. Therefore, using explicit feedback as a label for evaluating relevance ranking tasks is inappropriate. 
To this end, we defined user interests based on CWT in Eq.~\eqref{eq: interest define}. However, the unbiasedness of this interest label still needs to be discussed. 

To achieve duration unbiased, the user interest indicator $r_{u,v}$ needs to fulfill two characteristics: (1) it should be independent of video duration, and (2) when $r_{u,v}=1$, user explicit feedback should be equivalent across all video duration (thus mitigating the issue described in Fig~\ref{fig: behavior confusion}). To verify whether our interest indicator defined in Eq.~\eqref{eq: interest define} satisfies the above characteristics, we calculate the proportion of user interest labels $r_{u,v}$ and the proportion of explicit feedback when $r_{u,v}=1$ grouped by video duration.
Note that since videos shorter than $w_{0.7}$ cannot achieve a watch time of $w_{0.7}$, we exclude these videos from our analysis. The results are presented in Fig~\ref{fig: long_view verify}. It is evident that our defined $r_{u,v}$ roughly satisfies the above two characteristics on both the KuaiRand and WeChat datasets, thus indicating its unbiasedness.

\section{More Experimental results}
\label{appendix: more experimental results}

\subsection{Parameter sensitivity}
\begin{figure}
    \subfigure[watch time prediction ]{
    \includegraphics[width=0.22\textwidth]{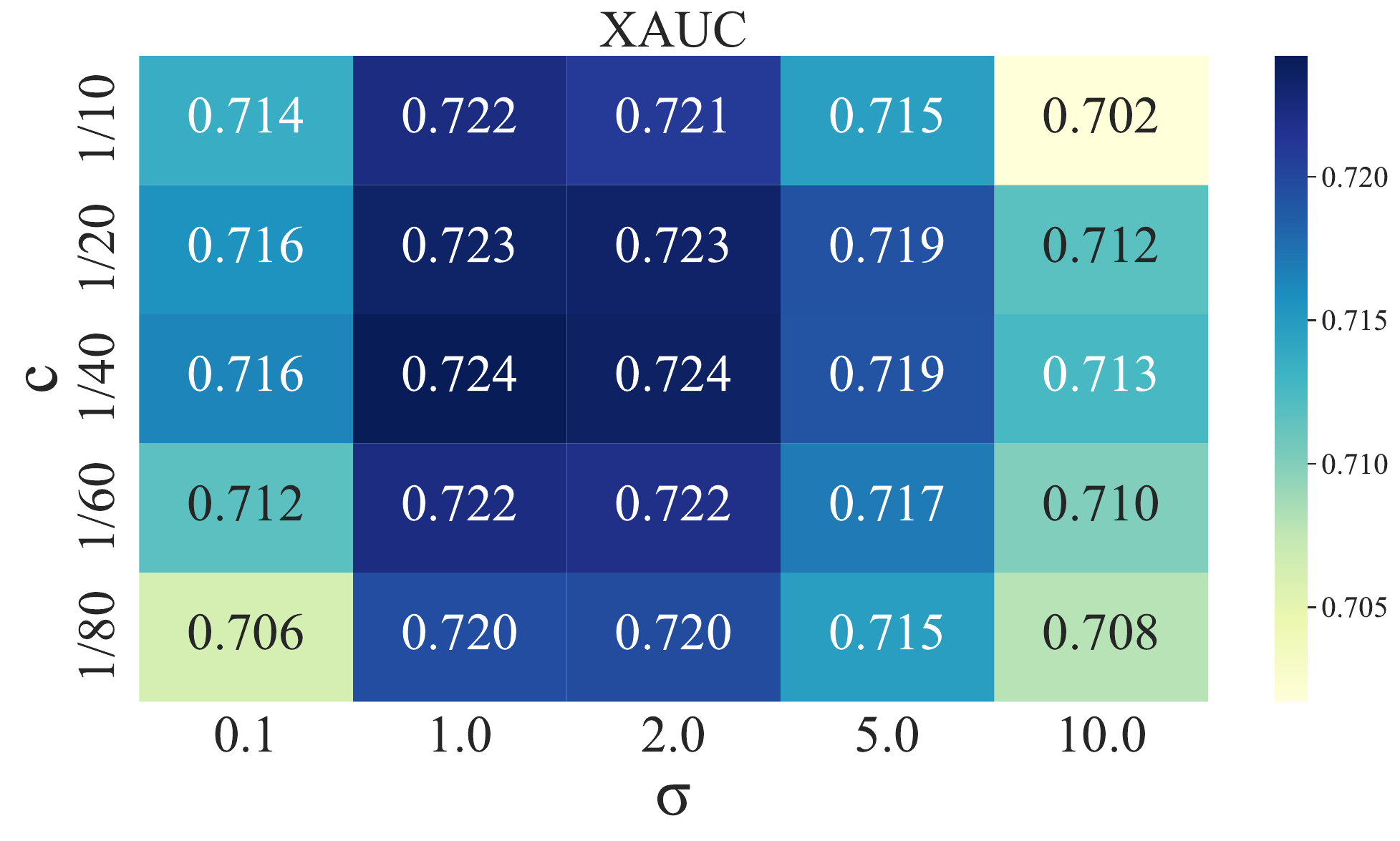}}
    \quad
    \subfigure[relevance ranking]{
    \includegraphics[width=0.22\textwidth]{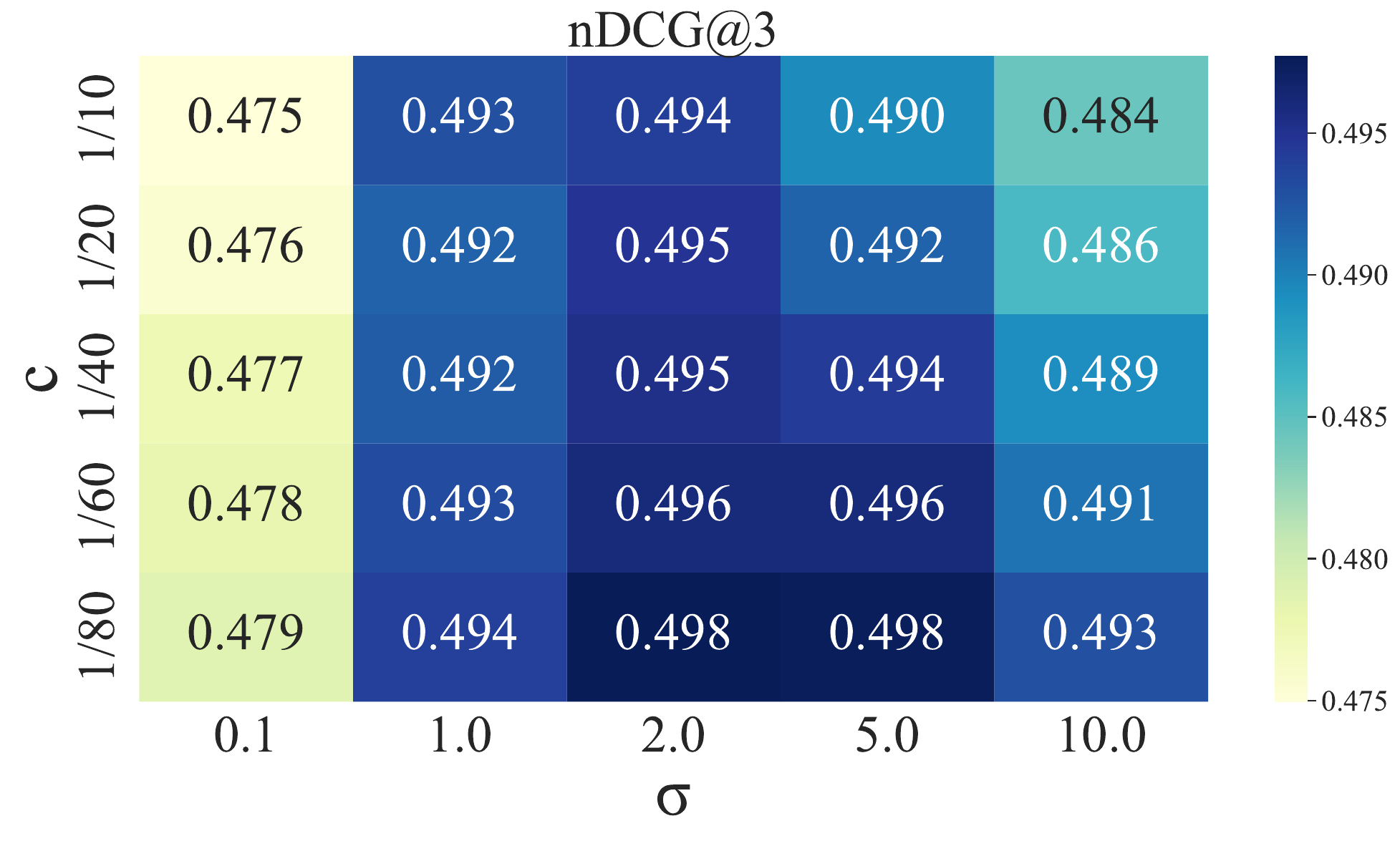}}
    \caption{The parameter sensitivity of CWM on KuaiRand dataset where the backbone model is FM.}
    \label{fig: hyper_param_sense}
\end{figure}

There are two hyper-parameters in the proposed CWM: one is the user watch cost $c$ in the cost-based transform function (Eq.~\eqref{eq: correct func inverse}). The larger the $c$, the more sensitive users are to watch time; Another is the variance term $\sigma$ of user interest in counterfactual likelihood function (Eq~\eqref{eq: likelihood parameterize}). The larger the value of $\sigma$, the more dispersed the user's interest distribution is. Fig.~\ref{fig: hyper_param_sense} illustrates the performance changes of recommendation with different values of $c$ and $\sigma$. For watch time prediction (Fig.~\ref{fig: hyper_param_sense}(a)), the best hyper-parameter is $\sigma\in(1.0,2.0) \land c\in(1/40,1/20)$; For relevance ranking (Fig.~\ref{fig: hyper_param_sense}(b)), the best hyper-parameters is $\sigma\in(2.0,5.0) \land c\in(1/80,1/60)$. Note that the best hyper-parameters of two tasks may not be the same. In practice, it is necessary to adjust the hyper-parameters to make CWM perform best.

\begin{figure}
    \subfigure[Duration$\leq$100s]{
    \includegraphics[width=0.22\textwidth]{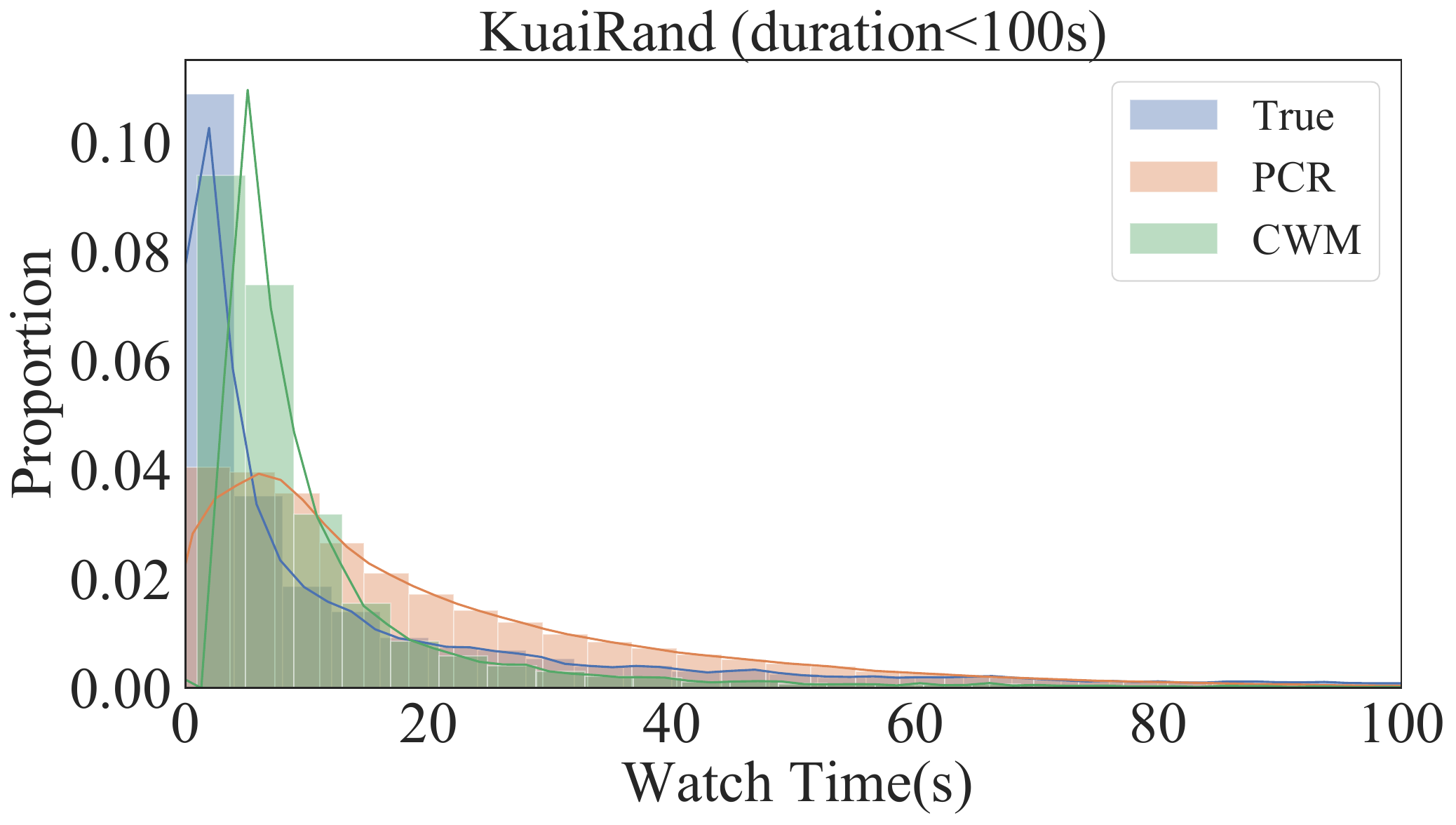}}
    \quad
    \subfigure[Duration=30s]{
    \includegraphics[width=0.22\textwidth]{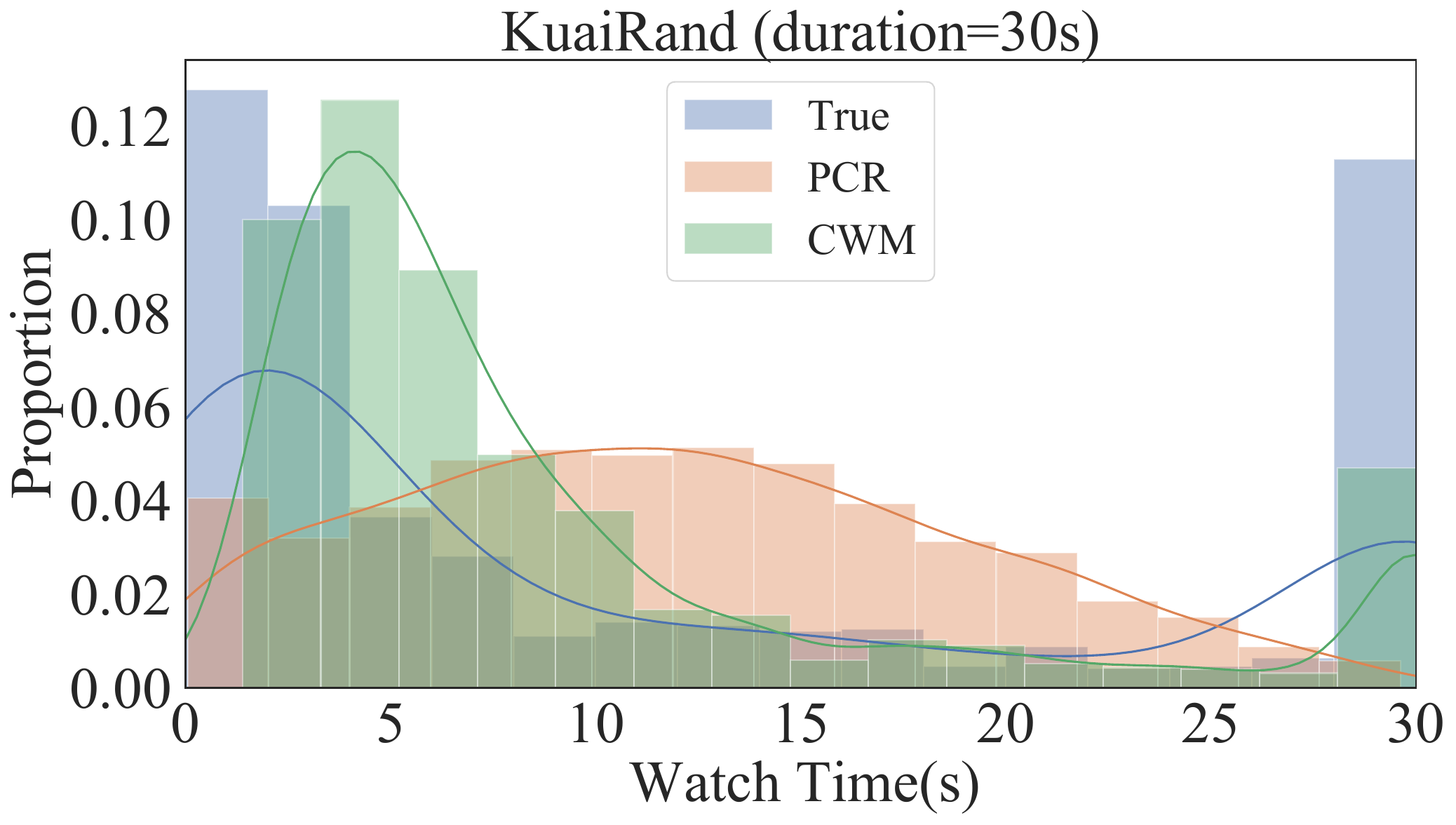}}
    \caption{The true and estimated watch time distribution on KuaiRand dataset.}
    \label{fig: case study}
\end{figure}

\subsection{Better fit to the true watch time distribution}
We examine whether CWM can fit the true watch time distribution. As a comparison, we also present the estimated watch time distribution by PCR, which is representative of existing debiasing methods. Fig~\ref{fig: case study}(a) shows the true and estimated watch time distribution on videos with less than 100s duration on KuaiRand. We can find that the estimated watch time distribution by PCR is more flattening than the true distribution. It overestimates higher watch time (i.e., >10s). In contrast, our CWM can fit the true distribution even better. Fig~\ref{fig: case study}(b) shows the true and estimated watch time distribution on videos with 30s duration on KuaiRand. It can be found that PCR only estimates a single-peaked distribution which differs significantly from the true bimodal distribution. In comparison, CWM can estimate a similar bimodal distribution to the true distribution, demonstrating CWM's effectiveness in watch time prediction.

\end{document}